\documentclass[pdflatex,sn-basic]{sn-jnl}
\usepackage{booktabs,array,graphicx,xcolor,amsmath,amssymb}
\graphicspath{{./figs/}}
\let\table\tableorg\let\endtable\endtableorg

\begin{document}
\title[Re-entering Spatial Prediction Contests]{Re-entering Thirty Years of Spatial Prediction Contests and Designing the Next}
\author*[1]{\fnm{Stefano} \sur{Castruccio}}\email{scastruc@nd.edu}
\affil*[1]{\orgdiv{Department of Applied and Computational Mathematics and Statistics}, \orgname{University of Notre Dame},
\orgaddress{\city{Notre Dame}, \state{Indiana}, \country{USA}}}

\abstract{Blind prediction competitions have been a cornerstone of model development in spatial statistics. Yet each competition
tested only a subset of established methods, and machine learning has since produced deep,
generative and pretrained models. Would a wider range of models, some classical methods the teams did not enter and new ones,
change the leaderboards? We re-enter five competitions with twenty-five methods, using the original data, test sets and scores.
The leaderboards would largely stand. Our best entry scores below the original winner on six of the twenty-nine,
but only slightly, and by a resolved margin only once. On two non-Gaussian fields an untransformed Gaussian process leads.
Methods differ mainly far from the observations, where predictions rest on assumptions. This suggests that what
matters is how well these assumptions match the data at the scales the sampling design probes, not a method's recency. We outline a new kind of competition, on data that strain classical assumptions, scored on
full predictive distributions and run in a standardized environment, which agentic AI workflows make practical; a pilot
shows that releasing archives of past fields lets generative models move beyond kriging.}


\maketitle

\section{Introduction}\label{sec:intro}
Blind prediction competitions are a recognized engine of methodological progress, from forecasting \citep{makridakis2020m4} to machine learning, where much of the field has advanced through shared datasets scored on hidden test sets
(the common task framework, \citealp{donoho2017}). The Netflix Prize offered US\$1
million to the first team that improved the RMSE of the company's movie-rating predictions by 10\%
\citep{bennett2007netflix}, and in the ImageNet challenge a deep convolutional network won by a wide margin in
2012, a result that led to the widespread adoption of deep networks in computer vision \citep{russakovsky2015imagenet}. The M4 forecasting competition, on 100{,}000
time series, showed the value of training one model across many series rather than one series at a time
\citep{makridakis2020m4}. This idea made it possible to forecast entirely new time series (zero-shot forecasting; \citealp{oreshkin2021zeroshot}) and led to
models pretrained on large collections of series, today's foundation models for time series \citep{ansari2024chronos}.

Spatial statistics has its own tradition of
such competitions. \citet{englund1990} gave the same data to twelve
geostatisticians and found a ``variance of geostatisticians'' as large as the variance of the data. The Spatial
Interpolation Comparison of 1997 \citep[SIC97;][]{dubois1998,dubois2003} opened the exercise to anyone: participants
received daily rainfall at 100 Swiss stations and predicted 367 withheld ones. SIC2004 \citep{dubois2005} asked for
automatic algorithms able to map ambient radioactivity in routine conditions and in a simulated emergency.
\citet{heaton2019} compared twelve methods for large spatial data on a satellite land-surface temperature field; their aim
was to compare ways of fitting one prescribed model (an exponential covariance with a linear trend in the coordinates). The
KAUST competitions \citep{huang2021,abdulah2022} compared approximations of Gaussian process inference on tens of thousands
of simulated locations. Comparative studies outside formal competitions find no method
uniformly best \citep{li2011review,heuvelink2022}. Each competition produced a leaderboard that reflected the state of the art of its time, but it tested only a subset of
established methods: no team in the second KAUST competition, for instance, entered kriging with external drift.

The methodological landscape has since changed. Random forests with spatial features \citep{hengl2018}, neural networks on
coordinates or radial basis expansions \citep{chen2024deepkriging}, graph neural networks \citep{kipf2017}, diffusion and
flow-matching generative models \citep{ho2020,lipman2023} and pretrained foundation models for tables
\citep{hollmann2025,qu2026tabiclv2,priorlabs2026tabpfn35} and locations \citep{vivanco2023geoclip,klemmer2023satclip} are now routinely proposed for
spatial prediction \citep{wikle2023,nag2023}. These methods are usually demonstrated on new datasets, often large and with complex,
nonstationary structure, where a flexible model has the advantage. They are rarely, if ever, tested on the datasets of past
competitions, on which the field has already compared its methods.

Suppose, then, that the spatial prediction competitions of the last thirty years were held again in 2026, with the same
data, the same test sets and the same scores, and with a wider range of models: the new methods, and some of the classical methods the
original teams did not enter. Would the leaderboards look different? To answer this question, we re-enter five of the best-known spatial statistics competitions. We run twenty-five methods from seven families,
from inverse distance weighting and kriging to tabular foundation models and diffusion models, using only the
information the original participants had; we score them on the original test sets with the organizers' metrics and
place them on the published leaderboards. None of the
methods is new: all have been published, and each enters with the configuration
selected on the training data only, without access to the test data. Running every method with one implementation, under one protocol and in one standardized computing
environment, on every leaderboard took about 300{,}000 CPU core-hours and 1{,}250 GPU-hours on a shared cluster, a task we
could achieve only with an agentic AI workflow. The same workflow built a repository that researchers and AI agents
alike can use to recompute every number in the paper, from a machine-readable list of the paper's claims and the command
that checks each one: a study and a standard of reproducibility far beyond what a single researcher could reach by hand.

In short, the leaderboards would barely change. Our best entry scores below the original winner on six of the
twenty-nine leaderboards of Table~\ref{tab:comp}, but only slightly: on four simulated fields, where the winners are very
close to the optimum, the margins are ties under our rules; on the satellite data, where a graph neural network leads, by an unresolved margin; and only on one simulated field by a barely resolved margin, there by kriging with external drift, a
classical method the teams did not enter. The contribution of the study is not a new
ranking of methods but what such a re-entry teaches about spatial prediction and about competitions, and we draw three
implications. First, on the competitions we re-entered, where most test locations lie near observations, kriging and
Gaussian processes remain the reference against which a new method should be compared: they give our best entry on
twenty of the twenty-nine leaderboards and are close to the winner on the others. Second, performance appears to depend on how well a method's assumptions match the
data, not on how recent the method is: where the field is stationary and well described by a parametric covariance, as
kriging assumes, flexible models have little to gain, and near the data almost any method does well. Third, these
datasets cannot separate the leading methods: where the optimal prediction is known, the winners are at or close
to it, except on one bivariate field, and on the real fields the leading entries differ by less than the sampling variability of one test set. New
competitions on more challenging data, where the classical assumptions of geostatistics (a stationary, isotropic, Gaussian
field on a Euclidean domain) fail, can change the picture. We show this with a pilot competition that we designed and ran
on daily precipitation over the Alps: with a decade or more of past fields released, a diffusion model predicted the
distribution of the rain better than kriging, and its accuracy on the mean improved steadily with the length of the
archive until, with 30 years, it was ahead of kriging.

Section~\ref{sec:comp} describes the competitions, Section~\ref{sec:methods} the methods and the protocol,
Section~\ref{sec:results} the results and Section~\ref{sec:discussion} their interpretation and the design of future
competitions. The full leaderboards, the tuning, the uncertainty of the scores, the calibration of the predictive
intervals and the implementation details are in the Supporting Information.

\section{The competitions}\label{sec:comp}
\begin{figure*}[!t]\centering
\includegraphics[width=\textwidth]{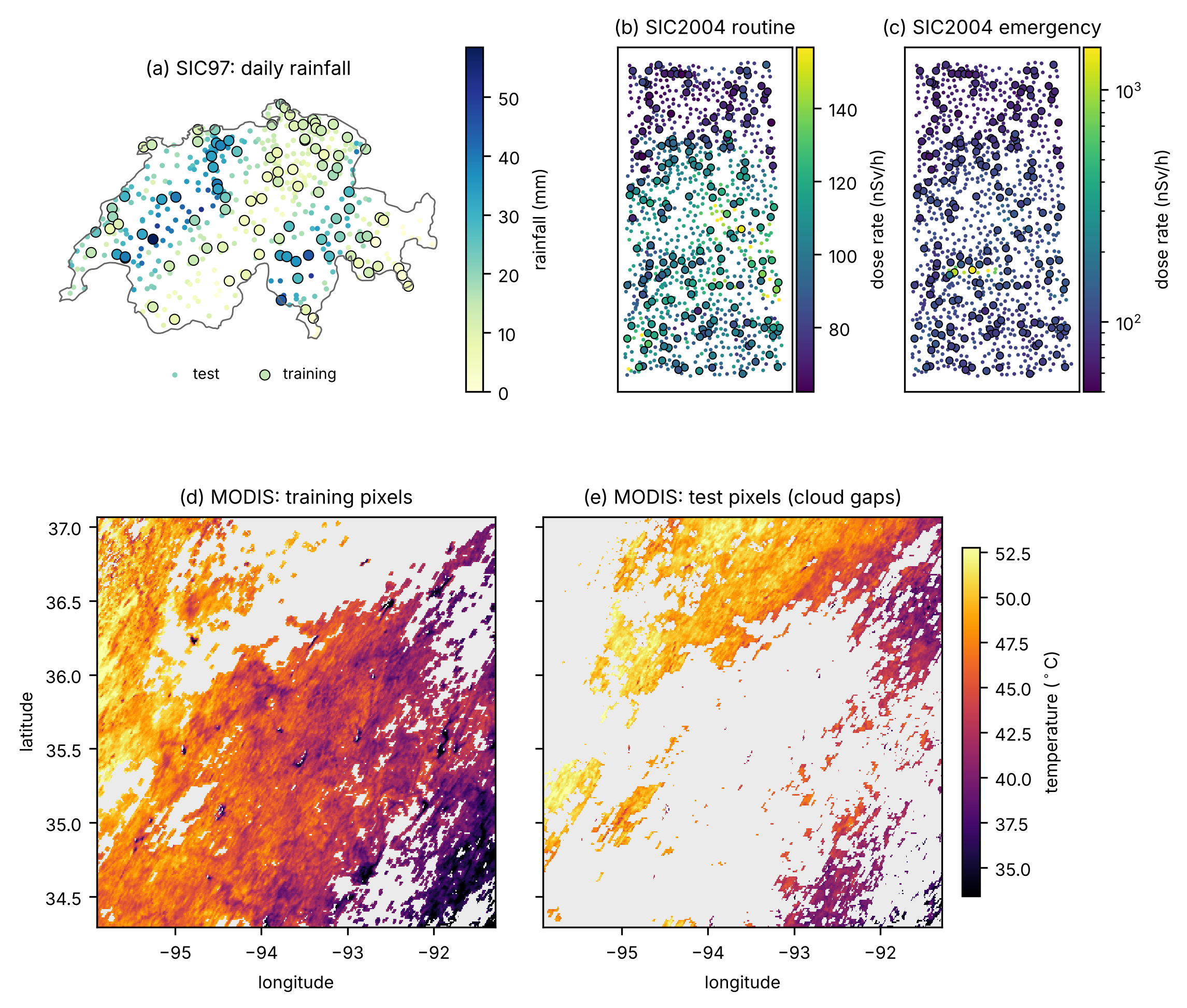}
\caption{The real-data competitions, colored by the observed value. In the top panels (a--c), the larger circles with
a black edge are the training stations, whose values the participants received, and the small dots are the test
stations, whose values were withheld. (a) SIC97, daily rainfall on 8 May 1986 at 100 training and 367 test stations in Switzerland. (b, c) SIC2004,
gamma dose rate at 200 training and 808 test stations on the test day, in the routine and the emergency scenario
(the latter on a logarithmic scale), in the coordinates released by the organizers. (d, e) MODIS land-surface temperature on
4 August 2016 \citep{heaton2019}: the 105{,}569 training pixels and the 42{,}740 test pixels, hidden by the cloud
cover of another day; gray: pixels not in the set shown.\label{fig:data}}
\end{figure*}

The five competitions are SIC97, daily rainfall at Swiss stations \citep{dubois2003}; SIC2004, gamma dose rates at
German monitoring stations in a routine and an emergency scenario \citep{dubois2005}; MODIS, a satellite land-surface
temperature field, with a field simulated on the same grid \citep{heaton2019}; and the first and second KAUST
competitions, KAUST1 and KAUST2, on simulated fields \citep{huang2021,abdulah2022}. Of KAUST1 we re-enter the sixteen
stationary Mat\'ern fields (G1--G16) and the two non-Gaussian fields (NG1, NG2); of KAUST2, the fields with a spatially
varying mean (1a-1) or covariance (1a-2) and the two bivariate fields (3a-1, 3a-2, variables $Z_1$ and $Z_2$), whose
larger versions are in the Supporting Information (Section~S1). Table~\ref{tab:comp} summarizes the twenty-nine
leaderboards, twenty-five of them simulated, and Figure~\ref{fig:data} shows the data of the three real-data
competitions.

\begin{table*}[!t]\centering\footnotesize\setlength{\tabcolsep}{4pt}
\caption{The competitions re-entered. $n$: training and test locations; entries: number of original entries with
published scores (in KAUST1, submissions; a team could submit several). KAUST1, KAUST2: the first and second KAUST
competitions; MODIS: the comparison of \citet{heaton2019}.\label{tab:comp}}
\begin{tabular*}{\textwidth}{@{\extracolsep\fill}l>{\raggedright\arraybackslash}p{3.6cm}>{\raggedright\arraybackslash}p{3.4cm}rrlr@{}}\toprule
Competition & Field & Leaderboards & \multicolumn{1}{c}{$n$ train} & \multicolumn{1}{c}{$n$ test} & Score & Entries \\ \midrule
SIC97 (1997) & daily rainfall, Switzerland & 1 & 100 & 367 & RMSE & 15 \\
SIC2004 (2004) & gamma dose rate, Germany & 2 (routine, emergency) & 200 & 808 & MAE & 31 \\
MODIS (2019) & MODIS land-surface temperature & 1 & 105{,}569 & 42{,}740 & RMSE & 12 \\
MODIS simulated (2019) & Gaussian, exponential covariance & 1 & 105{,}569 & 44{,}431 & RMSE & 12 \\
KAUST1 (2021), 1b & simulated Mat\'ern & 16 & 90{,}000 & 10{,}000 & RMSE & 31 \\
KAUST1 (2021), 2a & simulated non-Gaussian (Tukey $g$-and-$h$) & 2 & 90{,}000 & 10{,}000 & RMSE & 23 \\
KAUST2 (2022), 1a & simulated, nonstationary & 2 (varying mean, covariance) & 90{,}000 & 10{,}000 & RMSE & 11 \\
KAUST2 (2022), 3a & simulated bivariate Mat\'ern & 4 (2 datasets $\times$ 2 variables) & 45{,}000 & 5{,}000 & RMSE & 6 \\
\bottomrule\end{tabular*}
\end{table*}

In three of the competitions the participants received information besides the coordinates: a digital elevation
model in SIC97; the measurements of ten earlier days at the same stations in SIC2004, on which they designed their
automatic algorithms before mapping the test day; and both variables of the bivariate field at the training locations
in KAUST2 3a.

Section~S1.1 gives further details; two features matter for what follows. First, the
geometry of the test set: in SIC97, SIC2004 and the KAUST competitions the test locations are interspersed with the
training locations, whereas on MODIS and its simulated version they form the large contiguous gaps left by the cloud
cover of another day (Figure~\ref{fig:data}d, e). Second, what is known about the best possible prediction: on the
simulated fields of KAUST1 1b and KAUST2 3a the organizers also report the prediction with the true model and
parameters, which is optimal in expectation, and on the simulated MODIS dataset the teams were told to fit the true
model.

\section{The 2026 methods and the protocol}\label{sec:methods}
We call \emph{method} a published approach to spatial prediction, such as kriging with external drift or a random
forest, and \emph{predictor} a version of a method that is tuned and scored as a separate competitor: most methods give
one predictor, and two give several, run on different inputs. We enter twenty-five methods, thirty-two predictors, in seven families
(Table~\ref{tab:methods}, whose labels are used in the text, tables and figures; implementation details in Section~S4).

\begin{table}[!t]\centering\footnotesize\setlength{\tabcolsep}{4pt}
\caption{The methods and their labels. Each gives one predictor, except the neural network (NN, NN-cov, Trend + NN: on the coordinates, with the covariates and on trend residuals) and DeepKriging (the same three inputs, with our coarser basis or the authors', DeepKriging-P). All are entered on the twenty-nine leaderboards, except deepspat (eighteen; it failed on the others), the three predictors with covariates (the seven leaderboards with a covariate) and GeoCLIP and SatCLIP, which need real longitudes and latitudes (MODIS and its simulated version).\label{tab:methods}}
\begin{tabular*}{0.8\textwidth}{@{\extracolsep\fill}lll@{}}\toprule
Family & Label & Method \\ \midrule
Baselines & IDW & inverse distance weighting \\
 & Linear trend & linear trend by least squares \\ \midrule
Kriging and GPs & Ordinary kriging & ordinary kriging \\
 & Regression kriging & least-squares trend, kriged residuals \\
 & KED & kriging with external drift \\
 & GP & Gaussian process, anisotropic Mat\'ern \\
 & Vecchia GP & Gaussian process, Vecchia approximation \\
 & deepspat & deep compositional spatial model (deepspat) \\ \midrule
Tree-based & Random forest & random forest \\
 & RFsp & random forest on distances to the data \\
 & RF + kriging & random forest, kriged residuals \\
 & Gradient boosting & gradient-boosted trees (LightGBM) \\ \midrule
Neural networks & NN & fully connected neural network (3 predictors) \\
 & DeepKriging & DeepKriging (6 predictors) \\
 & NN-GLS & network mean, nearest-neighbor GP covariance \\
 & Neural process & local transformer neural process \\ \midrule
Generative & Diffusion & conditional denoising diffusion \\
 & Flow matching & conditional flow matching \\ \midrule
Graph NNs & GCN & graph convolutional network \\
 & GCN-L & GCN with masked observed values as input \\ \midrule
Foundation models & TabPFN & TabPFN v2 \\
 & TabPFN-3.5 & TabPFN-3.5 \\
 & TabICL & TabICL v2 \\
 & GeoCLIP & GeoCLIP encoder + network \\
 & SatCLIP & SatCLIP encoder + network \\
\midrule
\multicolumn{3}{l}{Total: 25 methods, 32 predictors} \\ \bottomrule
\end{tabular*}
\end{table}

The seven families are described in Section~S4.1. \emph{Kriging and Gaussian processes} are the
geostatistical baseline: ordinary kriging, regression kriging and kriging with external drift (KED), with a variogram
fitted by weighted least squares \citep{cressie1993}; a Gaussian process (GP) with an anisotropic Mat\'ern covariance;
the Vecchia GP, which fits that covariance by Vecchia likelihood on all the training points
\citep{vecchia1988,katzfuss2021vecchia,guinness2018}; and the deep compositional spatial model (deepspat;
\citealp{zammitmangion2022}), a nonstationary Gaussian process that warps the domain with learned functions.
\emph{Tree ensembles} split on the coordinates and any covariate: the random forest \citep{breiman2001}, RFsp, which
uses the distances to the observation locations as features \citep{hengl2018}, and gradient boosting
\citep{ke2017lightgbm}. \emph{Neural networks} take the coordinates as inputs (NN) or a multi-resolution basis of them
(DeepKriging; \citealp{chen2024deepkriging}); NN-GLS \citep{zhan2025nngls} fits a network trend under a nearest-neighbor
Gaussian process covariance and kriges its residuals, and the neural process \citep{garnelo2018np} predicts each
location from its nearest observations. \emph{Generative models}, denoising diffusion and flow matching
\citep{ho2020,lipman2023}, learn the distribution of the value at a location given its inputs, and predict with the mean
of their draws. \emph{Graph neural networks} \citep{kipf2017} act on a graph that links each location to its nearest
neighbors; GCN-L also receives the observed values of the neighbors. \emph{Foundation models} are trained once, in
advance, on a vast collection of data and are not refitted: the tabular models TabPFN, TabPFN-3.5 and TabICL
\citep{hollmann2025,priorlabs2026tabpfn35,qu2026tabiclv2} take the training data of a new problem and the test locations
and return the predictions in a single pass, and the location encoders GeoCLIP and SatCLIP
\citep{vivanco2023geoclip,klemmer2023satclip} turn the coordinates into features for a network. On the large datasets,
methods that cannot use all the training points predict each location from its nearest training points, whose number is
tuned; for the foundation models this conditioning set is called the \emph{context}.

\paragraph*{Protocol.}\hspace{0.5em}Each predictor is fitted on the training data and applied to the test locations,
which are scored against the published truth with the organizers' metric: RMSE, except MAE for SIC2004, as ranked by its
organizers. Each of our entries is placed on the original leaderboard: if
three original entries scored better, it would have placed fourth. It is compared with the winner or, where the
organizers named none (SIC97, MODIS and its simulated version), with the best published entry.

Every predictor is run with its default settings, and every tuned predictor is scored on twenty configurations (the
default and nineteen random draws from a model-specific search space), except the tabular foundation models, whose search
space is a small grid over the number of ensemble members and the size of the context. Each
configuration is scored by five-fold cross-validation within the training data, which we call inner cross-validation:
the training locations are split into five folds, each fold is predicted from the other four, and the test data are
never used. The configuration with the best inner cross-validation score, the default included, is the predictor's
entry.

The information the participants received besides the coordinates (Section~\ref{sec:comp}) enters our predictors as
one optional covariate: the station elevation read from the elevation model in SIC97, the mean of
the ten earlier days at each station in SIC2004 (any two of the ten days are correlated at 0.94 or more, so their mean
loses little) and the other variable of the pair in KAUST2 3a. In SIC2004 and KAUST2 3a the covariate is
observed only at the training locations, so at a test location we use its kriging prediction from the training
locations. To train the predictors on the same kind of value, at each training location we also use a kriging
prediction, from the other training locations, rather than the observed value. This is redone for every fit, so that
in inner cross-validation the held-out locations are treated as test locations. Each predictor that can do without the
covariate, elevation included, uses it only if inner cross-validation favors it, as for any other setting.

The cross-validation folds replicate the test
geometry: among random folds, spatial blocks and, on MODIS, folds shaped like the released cloud mask, we use the design
whose held-out points are as far from the remaining data as the test points are from the training data
\citep{mila2022}. This choice uses
only the coordinates of the training and test locations, which the participants also had, and no observed values. The
rule selects random folds on every leaderboard except MODIS and its simulated version. There it selects folds shaped
like the clouds: each holds out the training pixels covered by a randomly shifted copy of the real cloud mask
(Section~S2.2).

\paragraph*{Uncertainty in the scores.}\hspace{0.5em}A different but similar test set would give somewhat different
scores. We measure this with a spatial block bootstrap of the test locations \citep{lahiri2003}, which resamples fifty
spatially compact blocks of test locations (Section~S2.1). The original predictions are not
public, so an entry is compared with a published score through the ratio of its resampled score to that score; the
central 95\% of these ratios is the \emph{marginal} interval, and the difference is \emph{resolved} if this interval
excludes one. Two of our entries are scored on the same resamples, which gives a much narrower \emph{paired} interval.
Our score \emph{ties} a published score if the two are equal at the published precision. On 1b and 3a the organizers
also report the score of the true model with its true parameters, which is optimal in expectation; a winner below it,
as on 1b and 3a-2, is there by chance, so we also count as a tie any score below the winner's by less than the
winner's own margin below the true model.

\section{Results}\label{sec:results}
\begin{figure*}[!t]\centering
\includegraphics[width=0.9\textwidth]{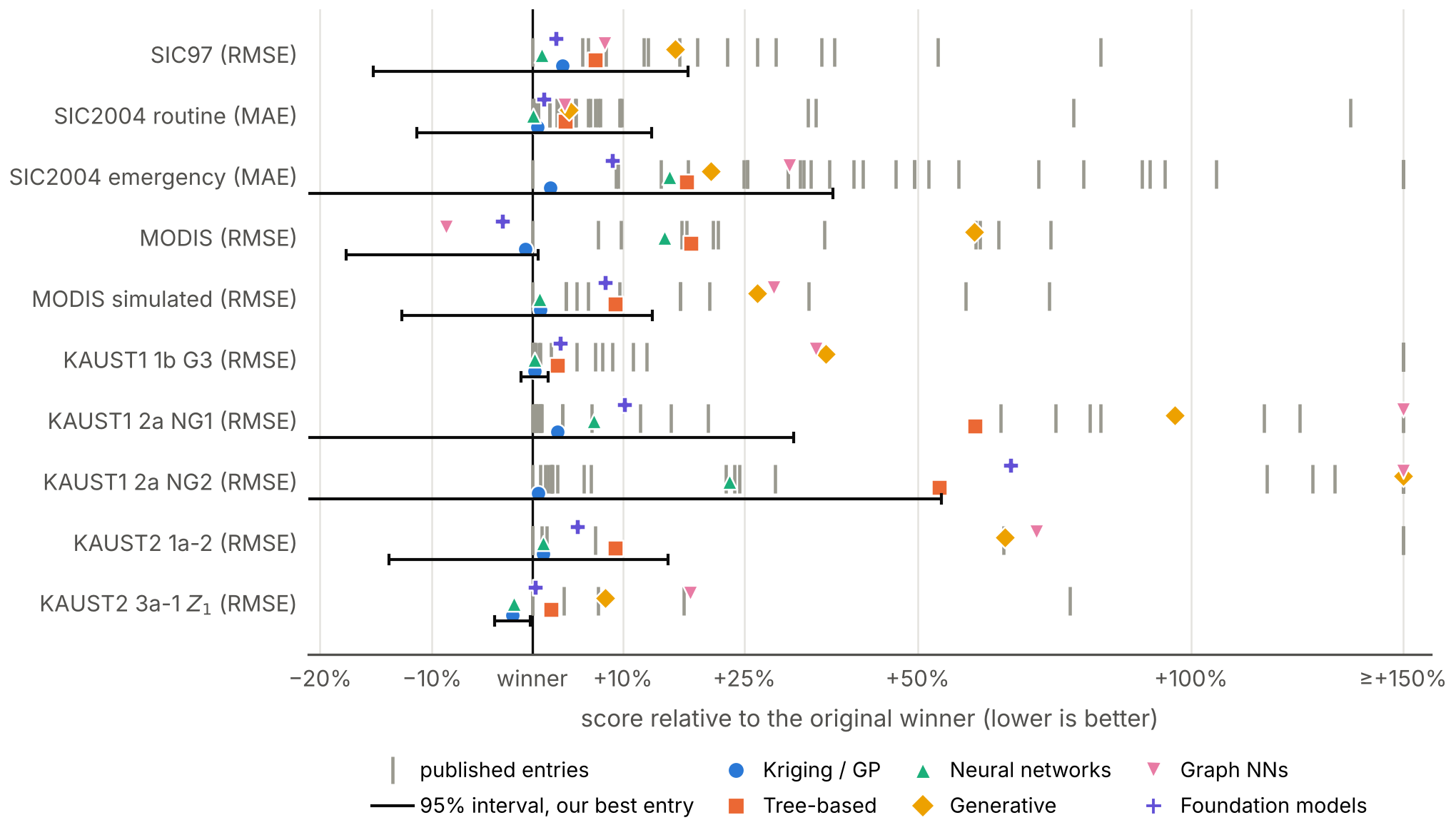}
\caption{Ten of the twenty-nine leaderboards, with the KAUST ones chosen as representatives (all twenty-nine are in
Figure~S1). Scores are relative to the original winner; values beyond $+150\%$ are drawn at the
edge. The bar under each row is the 95\% interval of our best entry against the winner: a margin is resolved only where
the bar lies entirely on one side of the winner's line, as on 3a-1 $Z_1$. Values below $-20\%$, and the parts of the bars
that extend past either limit, are also drawn at the edge.\label{fig:leaderboards}}
\end{figure*}

\subsection{The top of the leaderboards}
Figure~\ref{fig:leaderboards} places the best 2026 entry of each family, baselines aside, on ten representative leaderboards (all
twenty-nine in Figures~S1 and S2). A 2026 entry beats the best published entry by a
resolved margin on only one of them, the first
variable of the bivariate field 3a-1.

\textbf{3a-1, the only (barely) resolved margin.} Here every method that accepts a covariate may use the second
variable, and inner cross-validation keeps it for all the leading entries. KED beats the winner on the first variable by
2.0\% (95\% interval $[-3.9\%, -0.2\%]$); without the second variable, our best entry would place third. The margin
holds at the 95\% level but not at 99\%, nor with 25 rather than 50 bootstrap blocks (Section~S2.1);
it rests on the second variable, which was given to all predictors only after runs without it had been scored
(Section~S2.2); and the interval is not adjusted for the many comparisons we make, across
predictors and across leaderboards. NN-GLS scores $-1.9\%$ ($[-3.8\%, -0.1\%]$), and two more kriging entries are below
the winner by margins the bootstrap cannot resolve. All remain at least 4.1\% above the organizers' prediction
with the true bivariate model and parameters.

\textbf{MODIS, lower scores within the bootstrap noise.} Three 2026 entries have a lower RMSE than the best published
entry, the SPDE approach. GCN-L has the largest margin on any leaderboard with real data, $-8.7\%$, but it is not resolved: its interval, $[-17.8\%, +0.6\%]$, only just includes zero,
and whether it does depends on the number of bootstrap blocks (Section~S2.1). TabICL
($-3.1\%$, $[-13.7\%, +7.9\%]$) and KED ($-0.7\%$, $[-12.7\%, +11.8\%]$) are well within the sampling
noise, KED by little more than the last published digit. GCN-L is also below the SPDE approach in MAE, the other point score
\citet{heaton2019} report; TabICL and KED are not. The MODIS teams were asked
to use an exponential correlation where their method allowed it, while our entries chose their own covariance by
cross-validation. This freedom gives our entries an advantage the teams did not have, so small margins over the
winner on MODIS are, if anything, expected.

\textbf{3a-2, G14 and G16, negligible margins.} On the two variables of 3a-2 and on G16, the original winners already
score slightly below the organizers' prediction with the true model and parameters, which no method should beat except by chance. Our best entries score below them by even less, so under our
rules (Section~\ref{sec:methods}) they tie
with the original winners rather than beat them. On G14, the original winner is just above that prediction, and
regression kriging is below it by a margin that vanishes at the published precision, so it ties as well.

\textbf{The other twenty-three leaderboards, second place at best.} In SIC97, NN would have placed second; a post-competition
mixture of four machine-learning models, which was not a competition entry, reached a lower error \citep{gilardi2003}. In
SIC2004, NN-GLS would have placed second in the routine scenario, tied at the published precision, and the Vecchia GP
second in the emergency. On the simulated MODIS dataset, NN-GLS would have placed second. On the other KAUST2
leaderboards, the Vecchia GP would have placed third on the second variable of 3a-1, KED seventh on 1a-2 and flow
matching eighth on 1a-1. On the larger versions of 1a and 3a (sub-competitions 1b and 3b, with ten times as many training
locations), where we entered every predictor at its default configuration, our best entry is within 1\% of the original
winner on seven of the eight leaderboards, and TabPFN-3.5 scores below it on two by margins the bootstrap does not resolve
(Section~S1). On KAUST1, apart from the ties on G14 and G16, our best entry would have placed between second (G2, G10 and NG2) and 16th of 32 (G3), among
submissions separated by less than the sampling noise. On all but one, our best entry is between 0.004\% and 2.7\% behind the
winner, and every 95\% interval covers zero (Figure~\ref{fig:leaderboards}; Figure~S5). The
exception is G4, the smoothest field without a nugget, where the Vecchia GP is 3.7\% behind the winner
($[+0.1\%, +7.5\%]$): the only one of the twenty-nine leaderboards on which a published entry is ahead of every 2026 entry by a resolved margin.

\textbf{Kriging and Gaussian processes, best on 20/29 leaderboards.} Kriging or Gaussian process models are our best
entry on twenty of the twenty-nine leaderboards: nineteen simulated ones and the SIC2004 emergency. On the other three
with real data, our best entries come from other families. Where another family leads, kriging and Gaussian processes are still close: their best entry is at
most 3.7\% behind the winner on every leaderboard.

\textbf{One team, one win.} All of these placements treat each of our thirty-two predictors as a separate competitor. A single team entering the
competition would instead have submitted one method, chosen before seeing any test score. Had we entered as one team,
ranked our methods by their cross-validation score on the training data and submitted only the top one (Figure~S2), we would have won only on the first variable of 3a-1, where that method is NN-GLS
($-1.9\%$, $[-3.8\%, -0.1\%]$). On MODIS, regression kriging would have placed third of 13
(winner: the SPDE approach; $+8.0\%$, $[-10.4\%, +25.8\%]$), and between second and 25th on the other leaderboards.

\textbf{Calibrating each model matters.} Tuning can improve or degrade the test score, because cross-validation on
the training data is not the test set. On MODIS it moves KED from eighth with its default settings ($+23.9\%$) to first
($-0.7\%$; Section~S2), but it moves NN-GLS from second to fourth
($+2.8\%$ to $+14.9\%$). Overall, however, tuning helps: across the 768 combinations of
tuned predictor and leaderboard it lowers the test score in 85\% of cases, with a
median change of $-2.8\%$; where it raises the score, the median loss is 0.5\%.

\textbf{Cross-validation folds that mimic the test set.} The design of the cross-validation folds changes which configuration is selected, and with
it the test score: in an earlier, not fully comparable set of runs with generic $k$-means blocks instead of the cloud-mask cross-validation folds, the test
RMSE of a tuned model differed by $-18\%$ to $+25\%$. It
matters even more for choosing between models: the rank correlation between the cross-validation score of the selected
configuration and the test score is high on MODIS with the cloud-mask folds and was negative with $k$-means blocks in an
earlier set of runs. It is low
on the small competitions, SIC97 and SIC2004, and close to one on KAUST2 (Section~S2). Even the cloud-mask folds remain imperfect in two ways.
First, they are optimistic: the cross-validation errors are a median 21\% below the test errors.
Second, they do not identify the best model: they rank regression kriging and ordinary kriging above GCN-L, TabICL and
KED, the three entries that are best on the test set, so a team choosing by cross-validation would have submitted
regression kriging and placed third.

\subsection{Leaderboards on simulated fields}\label{sec:simulated}
This section covers the twenty-four KAUST leaderboards, simulated fields whose generating model is known: stationary
Gaussian fields (1b and 3a), non-Gaussian ones (2a) and nonstationary ones (1a). The
simulated MODIS dataset is discussed with the real MODIS field in the next section. On a Gaussian field with known
covariance, kriging is the optimal predictor \citep{stein1999}, so kriging and Gaussian process approaches are
expected to perform well on the stationary fields.

\textbf{Stationary fields, 1b and 3a: little room for improvement.} These twenty leaderboards are simulated
stationary Mat\'ern fields with randomly placed test locations: the sixteen Gaussian fields of 1b and the two
variables of the two bivariate fields of 3a. A stationary Gaussian process is the correct model here, and both the
winners and our best entries are at or very close to the optimal prediction: on every 1b field except G4, the smoothest without a nugget, our best entry is
within 0.5\% of the winner and of the organizers' prediction with the true model. The other families, NN-GLS included,
come close only on the rougher fields and on those with a nugget, where the noise masks the differences between methods;
on the smooth fields without a nugget they are far behind. On three smooth fields with a nugget a flexible method cannot
be separated from kriging: Trend + DeepKriging on G12 and G13 and TabPFN-3.5 on G15. The bivariate fields of 3a show the same pattern:
the leading families are close on the first variable of each field, every other family is far behind kriging on the smooth
second variable of 3a-1, and on the second variable of 3a-2 Trend + DeepKriging is narrowly ahead of the GP (Figure~S5). Our ranks on these leaderboards mean little anyway, because the differences between the leading
entries are much smaller than the width of the bootstrap intervals. The only headroom left is on 3a-1, where the
organizers' true-model prediction is better than every entry, the teams' and ours, by at least 4.1\% and 5.6\% on the
first and second variables.

\textbf{Non-Gaussian fields, 2a: Gaussian processes still lead.} NG1 and NG2 are simulated with a moderate and a strong departure from
Gaussianity, the only fields among the twenty-nine leaderboards that are non-Gaussian by construction. Methods that make
no Gaussian assumption should gain here, but they do not. On both fields the leaders are Gaussian processes: on NG1 the winner,
RESSTE, first transforms the data toward Gaussianity and then fits a Gaussian process to them, and on NG2 the winner,
HMatrix, fits a Gaussian process to the untransformed data. Our best entry on both fields is also an untransformed Gaussian process, the Vecchia GP: tenth of 24 on
NG1 ($+2.7\%$, $[-21.5\%, +31.7\%]$) and second on NG2 ($+0.6\%$, $[-53.4\%, +53.8\%]$). On NG2 the Vecchia GP is ahead of every submission that transformed the data. The other families are behind on
both fields; even the closest, NN-GLS, is behind the Vecchia GP by a margin the paired bootstrap resolves.

\textbf{Nonstationary fields, 1a: generic methods fall just short.} On 1a-1, simulated with a spatially varying mean, our best
entry is flow matching ($+1.0\%$, $[-0.4\%, +2.4\%]$), indistinguishable from the best
Gaussian process. Its marginal interval covers zero, but the gap is probably real: paired intervals between our
entries on this field are only about $\pm0.5\%$ wide, so a paired comparison with the winner, which would need its
predictions, would most likely resolve it in favor of the leading teams, which modeled the varying mean explicitly. On 1a-2, simulated with a nonstationary covariance, local KED is our best
entry ($+1.1\%$, $[-14.0\%, +15.4\%]$), and five teams lie within 0.03\% of each other ahead of
it, again likely a real gap. KED adapts to the slowly varying covariance without modeling it, by kriging from the nearest observations with a locally
estimated trend; explicitly nonstationary models instead build the variation into the covariance
\citep{paciorek2006,fouedjio2017}. Which approach works better remains open here: the only explicitly
nonstationary model we entered, deepspat, failed to fit on both 1a fields (Section~S4).

\subsection{Leaderboards on real fields}\label{sec:real}
This section covers the four leaderboards with real data: SIC97, the two SIC2004 scenarios and MODIS. Here the
generating model is unknown, so there is no optimal prediction to compare with, and the question is what separates the
methods.

\textbf{SIC97 and routine SIC2004, close behind the winner.} On both leaderboards our best entry would have placed
second: NN on SIC97, and NN-GLS on routine SIC2004, tied at the published precision with the
second-placed entry. The top of both is
crowded. On SIC97, TabPFN-3.5, KED and ordinary kriging are also within a few percent of the winner, a radial basis
function interpolator. On routine SIC2004, most original submissions are within 10\% of each other, and every family has
entries in this group (Section~S1). None of these differences is resolved, and on SIC97 the
elevation changes the leading entries little.

\textbf{SIC2004 emergency, second place by opposite routes.} Our best entry is the Vecchia GP ($+1.9\%$,
$[-26.9\%, +37.2\%]$), followed by TabPFN ($+8.8\%$, $[-39.5\%, +75.7\%]$); both would have placed second, while the other
kriging and Gaussian process entries are far behind. The two leading entries reach their scores in different ways
(Figure~S3). TabPFN never predicts more than about a fifth of the dose rate measured at the seven withheld
stations inside the release: it shrinks toward the background, which the MAE rewards. The Vecchia GP spreads the release
seen at two training stations over their neighborhood and predicts values of the right order of magnitude at the seven,
but its score is very sensitive to how the covariance is estimated (Section~S1).

\begin{figure}[!t]\centering
\includegraphics[width=0.85\textwidth]{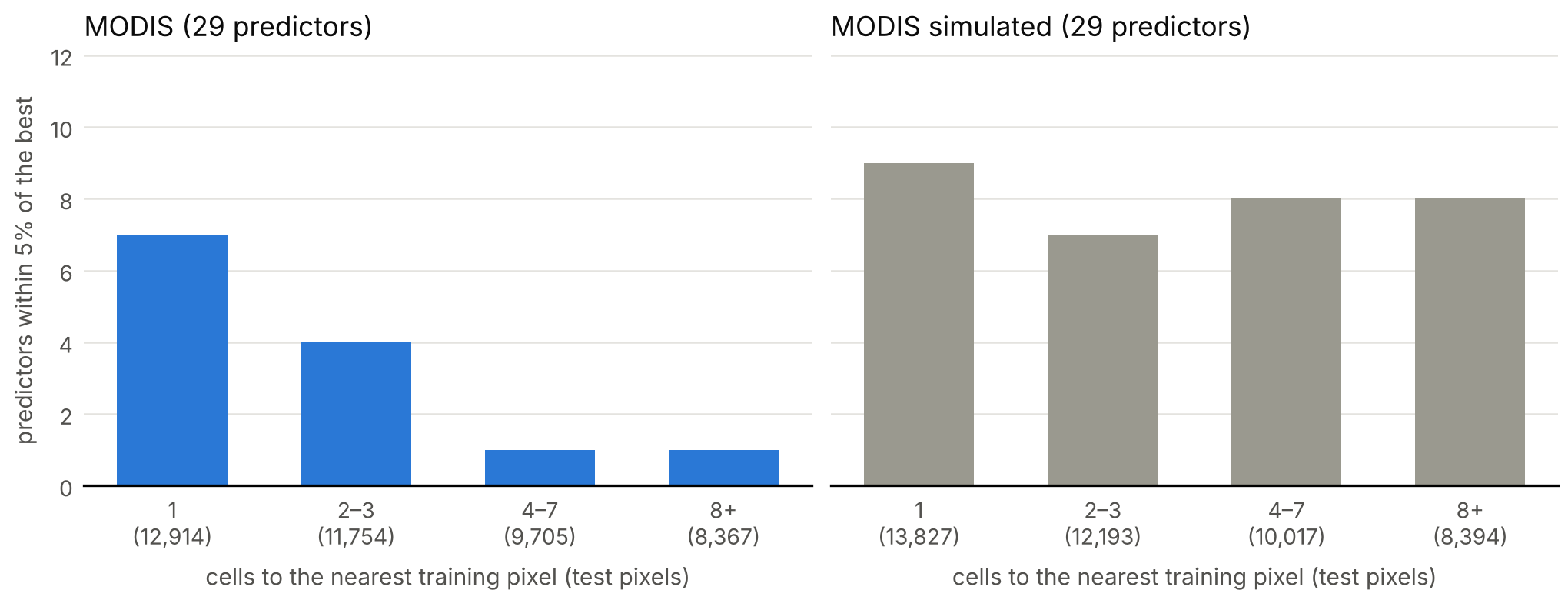}
\caption{Number of predictors whose test RMSE is within 5\% of that of the best predictor, with all test pixels
grouped by their distance to the nearest training pixel (in grid cells; number of test pixels in brackets), on MODIS (left) and on its simulated version (right), which has the same missing pixels but a field simulated
from a stationary Gaussian process. Near the data several predictors are as good as the best on both; eight or more cells
from the data, only one remains on the real field, against eight on the simulated one.}\label{fig:gapdist}
\end{figure}

\textbf{MODIS, the cloud gaps decide.} The test pixels lie at very different distances from the data: some at the
edge of a cloud gap, next to an observed pixel, others deep inside it (Figures~\ref{fig:data} and \ref{fig:gapdist}). Near the observed pixels prediction
is close to interpolation, and the predictors that use the nearby values perform alike: about a quarter of all
predictors are within 5\% of the best. Deep inside the gaps the prediction must extend the field over a large unobserved area, and the predictors
spread out: only one stays within 5\% of the best (Figure~\ref{fig:gapdist}). The hardest part of the test set is the hot northern gap, which the best entry of every family
underpredicts (Figure~S4). GCN-L,
TabICL and KED, the three entries below the winner, all predict each gap from the observations around it: GCN-L by
propagating them through its graph, TabICL by using the nearest training pixels as context, and KED by kriging from the
nearest observations, as did the best published entries in 2019, all Gaussian process models (the SPDE approach,
nearest-neighbor Gaussian processes, LatticeKrig; \citealp{lindgren2011spde,datta2016nngp}). GCN-L is worse than kriging
at the edge of the gaps but the best entry deep inside them, which is where it gains its lead over KED ($-8.0\%$, paired
interval $[-14.5\%, -0.3\%]$, a borderline margin: whether the interval excludes zero depends on the number of
bootstrap blocks; Section~S2).

\textbf{Simulated MODIS, the stationary pattern returns.} The simulated dataset keeps the grid, the training pixels and
the cloud gaps of MODIS, but replaces the temperatures with a stationary Gaussian field with the exponential covariance the
teams were told to fit. If the gaps alone shaped the results on MODIS, the same entries would lead here; if the field
matters, the pattern of the other simulated fields should return. It does: as on the KAUST fields, the methods that assume
a stationary covariance lead. NN-GLS, whose residual covariance is now the correct one, is our best entry and would have
placed second ($+0.8\%$, $[-12.8\%, +13.5\%]$), closely followed by the GP, the Vecchia GP and ordinary and regression
kriging, while GCN-L, TabICL and KED, our three best entries on MODIS, fall back. The predictors also do not spread
out deep inside the gaps: about as many are within 5\% of the best there as near the observed pixels
(Figure~\ref{fig:gapdist}). As on MODIS, a single test set cannot separate the leaders.

\textbf{Calibration and computing time.} Every predictor also gives a predictive distribution, and because all our
entries run in one standardized environment, their calibration and, except for the graph networks, which ran on GPUs,
their computing times can be compared directly (Section~S3). On MODIS no predictor is well calibrated: all underpredict the hot northern gap, and every
family's 90\% intervals under-cover there, whereas on the leaderboards whose test locations are interspersed with the
training data most families are close to the nominal level. Extra computing time does not buy accuracy or calibration:
KED and TabICL, among the most accurate entries on MODIS, take one to two minutes, while the slowest predictors take
hours without being more accurate or better calibrated.

\section{Discussion}\label{sec:discussion}
\paragraph*{What has not changed.}\hspace{0.5em}The top of the leaderboards: on twenty-eight of the twenty-nine no 2026 entry beats the
best published entry by a margin our bootstrap intervals can resolve. The best entry of SIC97 and the winners of SIC2004 keep first place in
the scores their organizers ranked by (RMSE in 1997, MAE in 2004); the winners of the two KAUST competitions,
mostly approximations of Gaussian-process inference designed for large datasets, are at or close to the optimal
prediction wherever it is known except on the bivariate field 3a-1, and
on G4, the smoothest KAUST1 field, the winner is ahead of every 2026 entry by a resolved margin; and
the SPDE approach of 2019 is beaten on MODIS only by unresolved margins. The
one barely resolved win, on a simulated bivariate field, comes from a wider range of models rather than from a new paradigm: KED, a classical method absent from the original leaderboard, and NN-GLS, a network whose residuals are
kriged. A kriging or Gaussian process model tuned on the training data alone would have come close to the winner
on every leaderboard, and it remains the reference against which a new method should be judged.

\paragraph*{What has changed.}\hspace{0.5em}Less than it may seem. Generic methods with no spatial design, which treat the
coordinates as ordinary inputs, now come within a few percent of the winners on most leaderboards, a difference a
single test set cannot detect. But this is not new: generic machine-learning methods were already as close on SIC97 two decades ago
\citep{gilardi2003}, and the tops of the 1997 and 2004 leaderboards were already diverse, with radial basis functions,
smoothing splines and neural networks among the best entries. Generic methods are not close everywhere, however: on the five
smooth 1b fields without a nugget, on the two non-Gaussian fields of 2a and on the smooth second variable of 3a-1 they are
far behind the winners. Where the winner is already at the optimal prediction, almost any
reasonable method comes close to it, and the leaderboard says little about which method is better. What has changed is
mainly the range of off-the-shelf tools that reach this level, which has widened without displacing the winners. The
closest any new method comes to doing so is on MODIS, where a graph neural network, GCN-L, has the lowest RMSE and MAE, by a margin at the edge of what the test set can resolve,
gained deep inside the cloud gaps (Figure~\ref{fig:gapdist}).
Forecasting competitions show a similar pattern, with classical models combined with networks leading in some and
gradient-boosted trees in others \citep{makridakis2020m4,makridakis2022m5}: which methods lead depends on the data.

\paragraph*{Assumptions and the sampling design.}\hspace{0.5em}In these competitions, which family wins is not related to the
size of the data: kriging or a Gaussian process is our best entry on fields with 200 stations and with 90{,}000
locations, and NN on one with 100. The comparison is limited, however, because size is confounded with the type of
data: the small datasets are real and the large ones mostly simulated. Our interpretation is that what matters is the match between a
method's assumptions and the structure of the data at the scales the prediction task probes, and that the sampling
design sets those scales. Kriging is the best linear predictor when the covariance is correctly specified, and remains
nearly so under mild misspecification \citep{stein1999}. When a test location lies close to observed ones, the
prediction depends almost only on the nearest observations and on the covariance at short distances
\citep{stein2002screening}, so the global assumptions (stationarity, the form of the mean, Gaussianity) barely matter,
and methods that use the nearby values perform alike; far from the data, the prediction rests on those global
assumptions, and methods separate. MODIS shows both regimes (Figure~\ref{fig:gapdist}). More than half of the test pixels lie within three grid cells of the
nearest training pixel. There, four to seven predictors are within 5\% of the best, from kriging and the Vecchia GP to
NN-GLS, GCN-L and the tabular foundation models. Deeper in the cloud gaps only one is,
against eight on the simulated version, whose field is a stationary Gaussian process with the same gaps. On the real
field, the stationary constant-mean models (the Vecchia GP, ordinary kriging) degrade sharply far from the data, kriging
with a trend and TabICL less, and GCN-L least, although
near the data it is the least accurate of the leaders. The pixels more than seven cells inside a gap are only a fifth
of the total, but they account for about twice that share of the squared error. It is the structure of the real field, not the geometry of the gaps
alone, that separates the methods there, although a single test set can barely resolve the differences among the
leading entries.

On the KAUST fields the test locations are scattered among tens of thousands of training locations, so nearly all of
them are close to observed ones. A flexible method should have the advantage over a stationary Gaussian model on the nonstationary and non-Gaussian
fields, yet the flexible methods among our entries had none, most likely for this reason. On the two nonstationary
KAUST2 fields they are tied with kriging on 1a-1 and behind it on 1a-2, and on the non-Gaussian fields a
stationary Gaussian process is our best entry. With dense data, moreover, a non-Gaussian marginal mainly changes the
shape of the predictive distribution, which RMSE and MAE barely reward.

A second factor also matters: whether a method uses the observed values near each prediction location. Near the data,
the methods that do, such as kriging, the Gaussian processes and the foundation models with local context, predict
directly from the nearby observations. Those that
learn only a fixed function of the coordinates, such as the coordinate networks, the random forests without kriged
residuals and the generative models, do not, and must reconstruct the local detail from the coordinates alone.
The foundation models show this directly: on MODIS, TabICL predicts much better when its context is the training pixels
nearest each test pixel than when it is a random sample of them. These results cannot
separate the contribution of locality from that of the covariance model, since kriging uses both. GCN-L, which conditions on nearby values without a covariance
model and leads on MODIS, is partial evidence for locality; a clean test would need more such methods \citep{sekulic2020}, compared on
nonstationary and anisotropic fields.

\subsection*{Future competitions: archives, agentic workflows and zero-shot kriging}
\begin{figure*}[!t]\centering
\includegraphics[width=\textwidth]{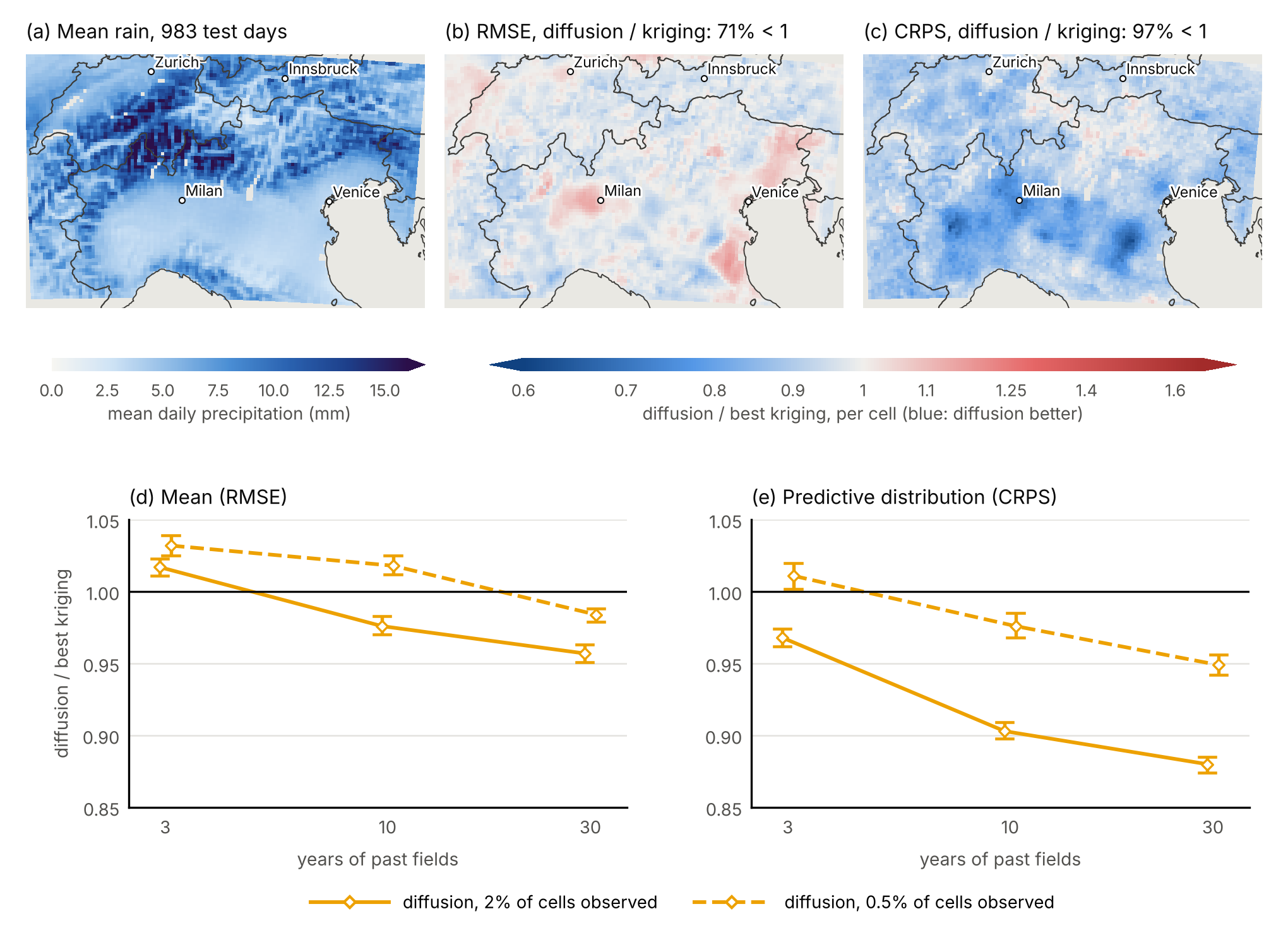}
\caption{Pilot competition on daily precipitation over the Alps (Section~S5). (a) Mean
precipitation over the 983 test days of 2021--2025. (b, c) Per-cell ratio of the RMSE and of the CRPS of the diffusion model
to those of the best kriging (archive kriging for the RMSE, censored-Gaussian kriging for the CRPS), with 10 years of past fields and 2\% of the cells observed (logarithmic scale; blue:
diffusion better). Titles give the share of land cells with a ratio below one. (d, e) The same ratios over all test days against the
length of the archive, with the 95\% intervals fixed in the design (test days resampled whole).\label{fig:cerra}}
\end{figure*}

Univariate prediction of stationary Gaussian fields is a mature benchmark: where the optimal prediction is known, the
winners are at or near it, and on the real fields the leading entries of thirty years cannot be separated by a single test
set. The competitions have saturated this problem, and twenty-five methods of 2026 change little at the top. The benchmark
remains useful for comparing accuracy per unit of computation at scale, the original aim of the KAUST competitions, but the
next competitions should pose harder problems. We offer seven recommendations. A pilot competition that we designed and ran for this purpose
(Figure~\ref{fig:cerra}; Section~S5) puts several of them into practice: daily precipitation, a skewed field with many
dry cells, released with up to 30 years of past fields, scored with proper scoring rules, and with every entry run in
one environment by an agentic workflow.

\emph{Data that challenge classical assumptions.} Competitions should move to settings where the assumptions of classical geostatistics are known to be
insufficient: multivariate fields with complex cross-dependence \citep{genton2015cross}; data on manifolds, such as global
fields on the sphere \citep{porcu2018} or stream networks \citep{verhoef2006}; nonstationary, non-Gaussian
and extreme-valued fields \citep{paciorek2006,huser2022extremes}; space--time forecasting; and the fusion of
heterogeneous sources with different supports \citep{gotway2002}. The
KAUST competitions already included multivariate and space--time tasks, and on 3a our best kriging or Gaussian process entry, which uses the other
variable at most as a covariate, beat or came within 0.2\% of the winning teams, so departing from classical assumptions is not enough by itself: the multivariate tasks should be designed so that the cross-dependence carries
information the marginal fields do not, for example with heterotopic sampling, in which the variables are observed at
different locations, or with cross-dependence that is nonlinear or varies in space.

\emph{Archives of past fields.} Many environmental fields come with decades of past replicates of the same field, and a
competition can release them with the training data. We tested this in a pilot competition on daily precipitation over
the Alps from the CERRA-Land reanalysis \citep{cerraland}, a field that is intermittent and far from Gaussian: on each
wet day the entrants saw a sparse network of cells and predicted the rest, with up to 30 years of past daily fields to
learn from: the entrants trained on the fields up to 2019, made their choices on 2020, and were scored on 2021--2025
(Section~S5). With
a decade or more of history, a diffusion model \citep{ho2020} predicted the distribution of the rain clearly
better than kriging, almost everywhere in the domain (Figure~\ref{fig:cerra}). On the mean, its advantage grew with
the length of the history. With 3 years of past fields diffusion was behind kriging with the covariance learned from those
fields. With 30 it was ahead at both densities of the observed network. Released history is what lets generative models move beyond
kriging, and the gain shows in the predictive distribution more than in the mean.

\emph{Test locations away from the data.} Where test locations lie among dense observations, nearly every method that
uses the nearby values comes close to the best prediction (Figure~\ref{fig:gapdist}). Test sets should place a substantial
share of the prediction locations far from the data, in large gaps, outside the sampled region or in the future, and
scores should be reported by distance to the nearest observation.

\emph{Scores that reward the whole predictive distribution.} \citet{heaton2019} already scored the continuous ranked probability score (CRPS), interval scores and
coverage, and SIC2004 reported several error measures; but rankings were still read from RMSE or MAE, which, as the SIC2004
emergency shows (Figure~S3), can reward shrinkage. Future competitions should rank on proper scoring rules, which compare the whole predictive distribution with the
observed value \citep{gneiting2007}. For spatial fields, they could also ask entrants for joint predictions
over all test locations, as a set of simulated fields, and score how well these reproduce the spatial dependence
\citep{gneiting2008energy,scheuerer2015variogram}, together with scores for the quantities that matter in applications, such
as exceedance probabilities in an emergency. In our pilot the two kinds of score disagree: with a decade of history and
the sparser network, diffusion predicted the distribution of the rain better than kriging, but its mean worse.

\emph{Designs with statistical power.} A single test set leaves most rankings within sampling noise: on SIC97, SIC2004 and
MODIS the score of a single entry varies by several percent from one bootstrap sample to another, and far more in the
SIC2004 emergency, so differences of that size between
entries cannot be detected against a published score, whereas the same differences are often detectable when two entries
are compared on the same test locations, on routine SIC2004 and the KAUST fields, only just on MODIS, though not on SIC97
or the SIC2004 emergency (Section~S2). Organizers should therefore release the
participants' predictions to allow paired tests \citep{hering2011}, use several independent realizations or test regions,
withhold enough test locations to detect the differences they care about, and, as the KAUST competitions did, report the
prediction with the true model, so that it is clear how far the entries are from the best possible prediction.

\emph{Standardized execution with agentic workflows.} Teams would still develop their code independently, and could submit several methods, but
the organizers would run every entry in a single standard environment, with the same data, hardware and computing
budget. Past competitions collected predictions that each team computed on its own systems; running dozens of
independently written codes in one environment would have been impractical by hand, but agentic AI workflows can now
install, run and check them almost automatically, as we did here for twenty-five methods. Because every entry would be
run within the same environment, anyone could reproduce its scores, and entries could be compared at the same computing
cost.

\emph{Towards zero-shot kriging.} In time-series forecasting, foundation models such as Chronos \citep{ansari2024chronos}
and TimesFM \citep{das2024timesfm} forecast series they have never seen, after pretraining on large and diverse
collections of real and synthetic series. The spatial analogue, a model that predicts the field of a new region without
that region's history, is the natural next challenge, and our pilot shows that it is not easy: diffusion models trained on other regions predicted the Alps' rain less
accurately than kriging fitted to the day's observations alone, and their predictive distributions were only modestly
better. TabICL fitted to the day alone,
which needs no archive, gave a better predictive distribution but not a better mean (Section~S5). The bottleneck is less
the volume of environmental data than its diversity: a time-series model sees many largely independent series, whereas a
spatial model sees, for each region, one field per day shaped by that region's orography and climate. Geostatistics can contribute here, with archives curated across many regions and with synthetic
fields drawn from spatial models, much as Chronos and TabPFN are pretrained partly on simulated data. Competitions that
withhold whole regions would measure the progress.

\section{Conclusions}\label{sec:conclusions}
With a wider range of models, classical and new, the leaderboards of five spatial prediction competitions held over
thirty years would largely stand. This says more about the competitions than about the methods we added, classical and
new: most of their test locations lie close to observed ones, where almost any method that uses the nearby values predicts
well, and methods separate only where the prediction must rest on their assumptions, far from the data. On six of the
twenty-nine leaderboards our best entry scores below the original winner, but on four simulated fields these are ties
(three at the optimum, one at the published precision), on the satellite data a graph neural network leads by an unresolved margin, and only on one
simulated field is the margin resolved, barely, with kriging with external drift, a classical method, and NN-GLS.

For practitioners, a kriging or Gaussian process model tuned with cross-validation folds that imitate the prediction task remains the
reference a new method has to beat on data like these; flexible methods are most useful where a stationary model cannot
represent the structure, and their uncertainty should be checked on designs that mimic the task
\citep{meyer2021aoa,mila2022}, or recalibrated \citep{mao2024conformal}, before it is used. For developers of
new methods, old leaderboards are a cheap and fair test, provided the comparison uses the information the original
teams had, selects configurations blind to the test data, and reports how far apart the scores really are. For
organizers, the next competitions should use data that challenge classical assumptions, release the history of the
field, place test locations away from the observations, and score full predictive distributions, so that they test the
new toolbox where it can make a difference; prediction in regions never seen in training remains an open challenge.

\bmhead{Supporting Information}
Supporting Information is available online: the competitions and their full leaderboards (Section~S1); the sampling
uncertainty of the scores, tuning, the sensitivity of the placements to how entries are chosen, including what we would
have achieved entering as one team, and the additions of the development phase (S2); calibration and, on MODIS, computing
time (S3); the methods and their implementation (S4); and the precipitation pilot competition (S5).

\section*{Declarations}

\bmhead{Data and code availability}
The competition data are publicly available from their original sources \citep{dubois2003,dubois2005,heaton2019,huang2021,abdulah2022};
the per-submission scores of the first KAUST competition are in the organizers' online assessment sheet. The data, code
and predictions of all reported predictors are in the repository
\url{https://github.com/stefanocastruccio/26.Castruccio.spatial-contests-reentry}, the larger prediction files as a download of its
release v1.0. The exception is sub-competitions 1b and 3b of the second KAUST competition (Supporting Information only):
their data are too large for the repository and are available from the organizers' release, and the repository holds
summaries of our predictions rather than the predictions themselves. The precipitation pilot is a second exception: the CERRA-Land fields are
available from the Copernicus Climate Data Store, and the repository holds the pilot's code, design documents and
scores, but not its per-day predictions or trained diffusion models. The repository includes guides for researchers and AI
agents, a machine-readable list of every quantitative claim with the command that recomputes it, and checksums for all the
data and predictions it holds. TabPFN v2 and TabPFN-3.5 are used under their Prior Labs licenses (the latter for non-commercial use
only); their weights, like those of SatCLIP, are not redistributed.

\bmhead{Use of artificial intelligence} Claude (Anthropic), used through the Claude Code agent, assisted the author in
running the computations, writing the code and drafting the text. The research question, the protocol, the interpretation
and the conclusions are the author's, who reviewed the code, results, citations and text and takes full responsibility
for the content. No AI tool was used to create or alter data. In accordance with Springer Nature's policy, the AI tool is
not an author.

\bmhead{Competing interests}
The author declares no competing interests.

\bmhead{Funding and acknowledgements}
This work was supported by the U.S. National Science Foundation under grants CDS\&E-MSS 2447624 and CISE OAC 2347239.
Computations were performed at the Center for Research Computing of the University of Notre Dame.

\bibliography{refs}
\end{document}


\maketitle

\section{The competitions and the full leaderboards}\label{sec:full}
\subsection{The competitions}\label{sec:comps}
\paragraph*{SIC97.}\hspace{0.5em}Participants mapped daily rainfall in Switzerland on 8 May 1986 from 100 stations and were scored on
367 withheld stations \citep{dubois1998,dubois2003}. Twenty methods were submitted, from kriging, splines and inverse
distance weighting to neural networks and fuzzy-logic interpolators; the organizers deliberately named no winner, and we
recovered the scores of fifteen entries from the papers reprinted in \citet{dubois2003}. Besides the station
coordinates, participants received a digital elevation model of Switzerland. Our entries use the elevation of each
station read from that model, so they have the same information as the participants.

\paragraph*{SIC2004.}\hspace{0.5em}Participants designed automatic mapping algorithms on ten days of gamma dose rates at 200 German
monitoring stations and then mapped, without human intervention, a new day from 200 stations onto 808 withheld ones
\citep{dubois2005,dubois2005intro}. Two versions of that day were released: a routine one and an emergency one, in which a simulated
release raised nine stations above 600\,nSv/h, seven of them among the withheld ones. Thirty-one submissions were scored by MAE, RMSE and correlation
\citep[Table 4]{dubois2005}. Participants therefore had, besides the 200 values of the test day, ten earlier days at the
same stations with which to design and calibrate their algorithms; the emergency winner, for example, used them to set
the range of its smoothing parameter \citep{timonin2005}. Our entries have the same information: the earlier days enter
as an optional covariate, their mean at each station, which each predictor may use or ignore, a choice made by
cross-validation on the training data (main text, Section~3).

\paragraph*{MODIS land-surface temperature.}\hspace{0.5em}\citet{heaton2019} compared large-data spatial methods on
daytime land-surface temperatures measured by the Terra MODIS instrument on 4 August 2016 over a $500\times300$ grid in
the south-central United States. Of the 148{,}309 pixels with a recorded temperature, 105{,}569 are for training and the
42{,}740 pixels hidden by the cloud cover of a different day form the test set, so the test pixels come in large contiguous
gaps. The only inputs are the coordinates; the location encoders (Section~\ref{sec:families}) use longitude and latitude. \citet{heaton2019} did not declare a winner; we call the entry with the lowest
RMSE (the SPDE approach of \citealp{lindgren2011spde,rue2009inla}) the best published entry; the other entries included
nearest-neighbor Gaussian processes \citep{datta2016nngp,finley2019nngp}, LatticeKrig \citep{nychka2015lattice} and the
nonparametric Gapfill method \citep{gerber2018gapfill}. The organizers asked the teams to use an exponential correlation and a nugget where their method allowed it, and the
teams used a linear trend in the coordinates, whereas our entries choose their own, so they compete
under a looser constraint than the original teams. \citet{heaton2019} also released a simulated dataset on the same grid,
with the same 105{,}569 training pixels: a Gaussian process with constant mean, exponential covariance and nugget, with
parameters estimated from the MODIS data, and every other pixel (44{,}431) in the test set. The teams were told to fit
that model, so on this leaderboard the prescribed model is the true one. We re-enter it as a separate leaderboard (the simulated MODIS dataset
below), with the same predictors and folds as MODIS.

\paragraph*{First KAUST competition.}\hspace{0.5em}The first KAUST Competition on Spatial Statistics for Large Datasets
(KAUST1 below; \citealp{huang2021}), held in 2020--2021, simulated fields on $[0,1]^2$ with the ExaGeoStat software
\citep{abdulah2018exageostat}, each with 100{,}000 locations split at random into 90{,}000 training and 10{,}000 test locations. Sub-competition 1a
asked for estimates of the Mat\'ern parameters of sixteen zero-mean Gaussian fields; 1b for predictions at the test
locations of the same sixteen fields; 2a for predictions on two non-Gaussian fields, obtained by applying a Tukey
$g$-and-$h$ transformation \citep{xu2017tukey} to a Gaussian Mat\'ern field, with a moderate (NG1) and a strong (NG2) departure from Gaussianity;
and 2b for predictions on larger datasets. Prediction entries were ranked by RMSE, and the organizers published the RMSE
of every submission on every dataset; teams could submit more than one method. We re-enter 2a and all sixteen 1b
fields (G1--G16 below), stationary Mat\'ern fields that differ in smoothness, range and whether they include a nugget
(parameters in Table~\ref{tab:kaust1}). We did not enter 1a, which concerns parameter estimation, nor, to limit
the computing budget, 2b. For 1b the organizers also report the prediction
with the true model and parameters.

\paragraph*{Second KAUST competition.}\hspace{0.5em}The second KAUST Competition on Spatial Statistics for Large Datasets
(KAUST2 below; \citealp{abdulah2022}) compared approximations of Gaussian process inference on fields simulated by
ExaGeoStat. We re-enter
its spatial prediction sub-competitions whose test values were released: 1a, two univariate fields on $[0,1]^2$ with a
spatially varying mean (1a-1) or a nonstationary covariance (1a-2); and 3a, three datasets, of which 3a-1 and 3a-2 can be scored, from the bivariate Mat\'ern model \citep{gneiting2010matern}, whose two variables
(labeled $Z_1$ and $Z_2$ in the tables and figures) are observed together at the training locations. The teams had both variables. Our entries predict each
variable with the other available as an optional covariate. Neither variable is observed at the test locations, so
there the covariate is the kriging prediction of the other variable from its training values. Our entries thus use the
same data as the teams, through a two-stage approximation to cokriging rather than a joint model. In the released solution file the test values of
3a-3 duplicate those of 3a-2, so 3a-3 cannot be scored. Sub-competitions 1b and 3b repeat 1a and 3a with ten times as
many training locations; their solution file holds distinct test values for all three 3b datasets, so all are scored,
and we entered them with every predictor at its default configuration only (KAUST2 1b and 3b below; Table~\ref{tab:kaustb}).
The other sub-competitions concern parameter estimation or space-time data. For 3a the organizers also report the prediction with the true model and parameters, which is optimal in
expectation and measures how much room for improvement is left.

\subsection{All leaderboards}
\begin{figure}[p]\centering
\includegraphics[width=\textwidth,height=0.82\textheight,keepaspectratio]{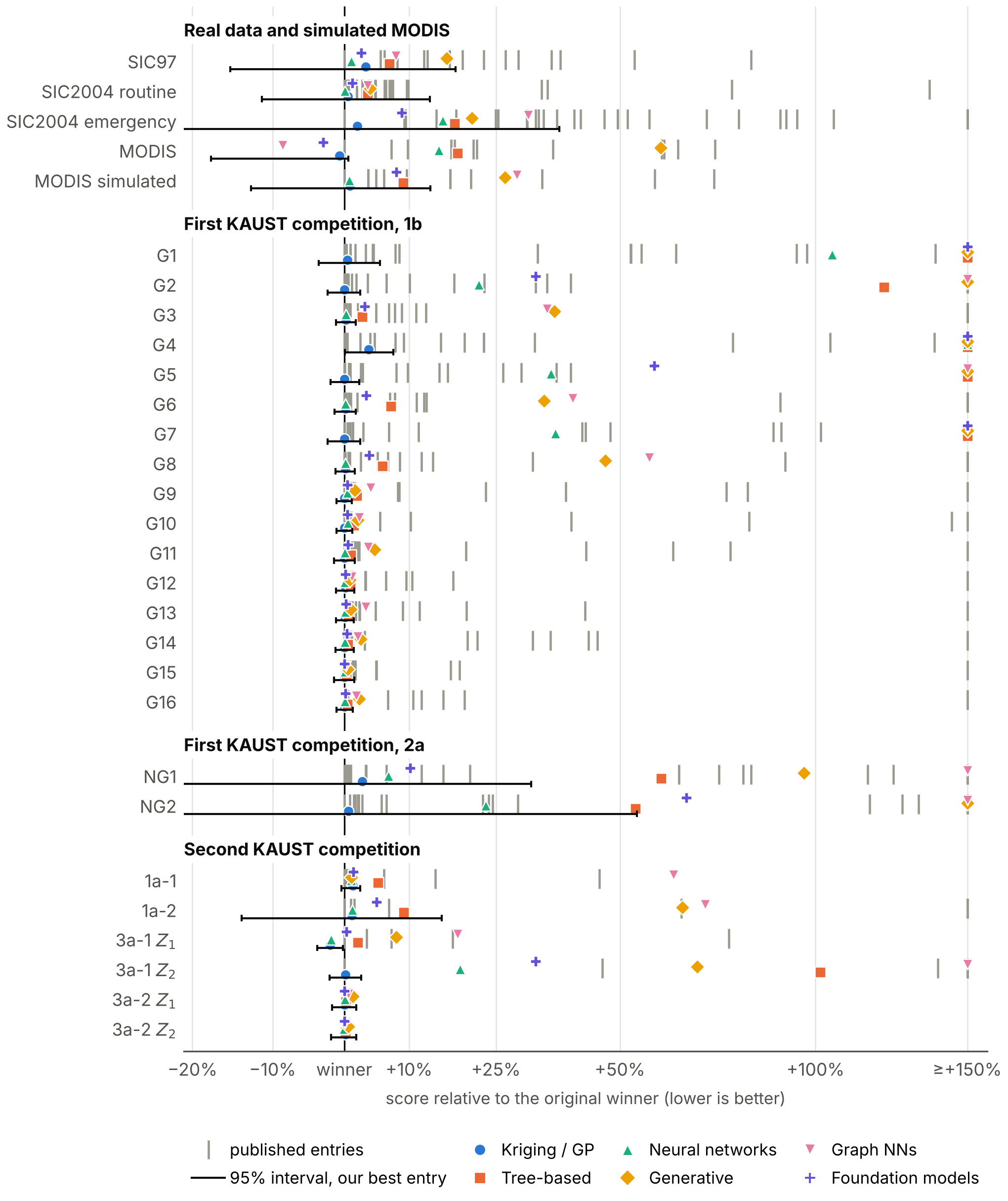}
\caption{All twenty-nine leaderboards re-entered (the main text shows ten); scores are RMSE, except MAE for SIC2004. Each
row is a leaderboard; gray ticks are the
published entries and colored markers the best 2026 entry of each family (baselines omitted), as the ratio of its score to that of the
original winner, on a logarithmic axis labeled in percent relative to the original winner; values and bars outside $-20\%$ to $+150\%$ are
drawn at the edges. The black bar under each row is the marginal 95\% block-bootstrap interval of our best entry against
the original winner's published score (Figure~\ref{fig:bootstrap}).\label{fig:leaderboards_all}}
\end{figure}

\begin{figure}[p]\centering
\includegraphics[width=\textwidth,height=0.7\textheight,keepaspectratio]{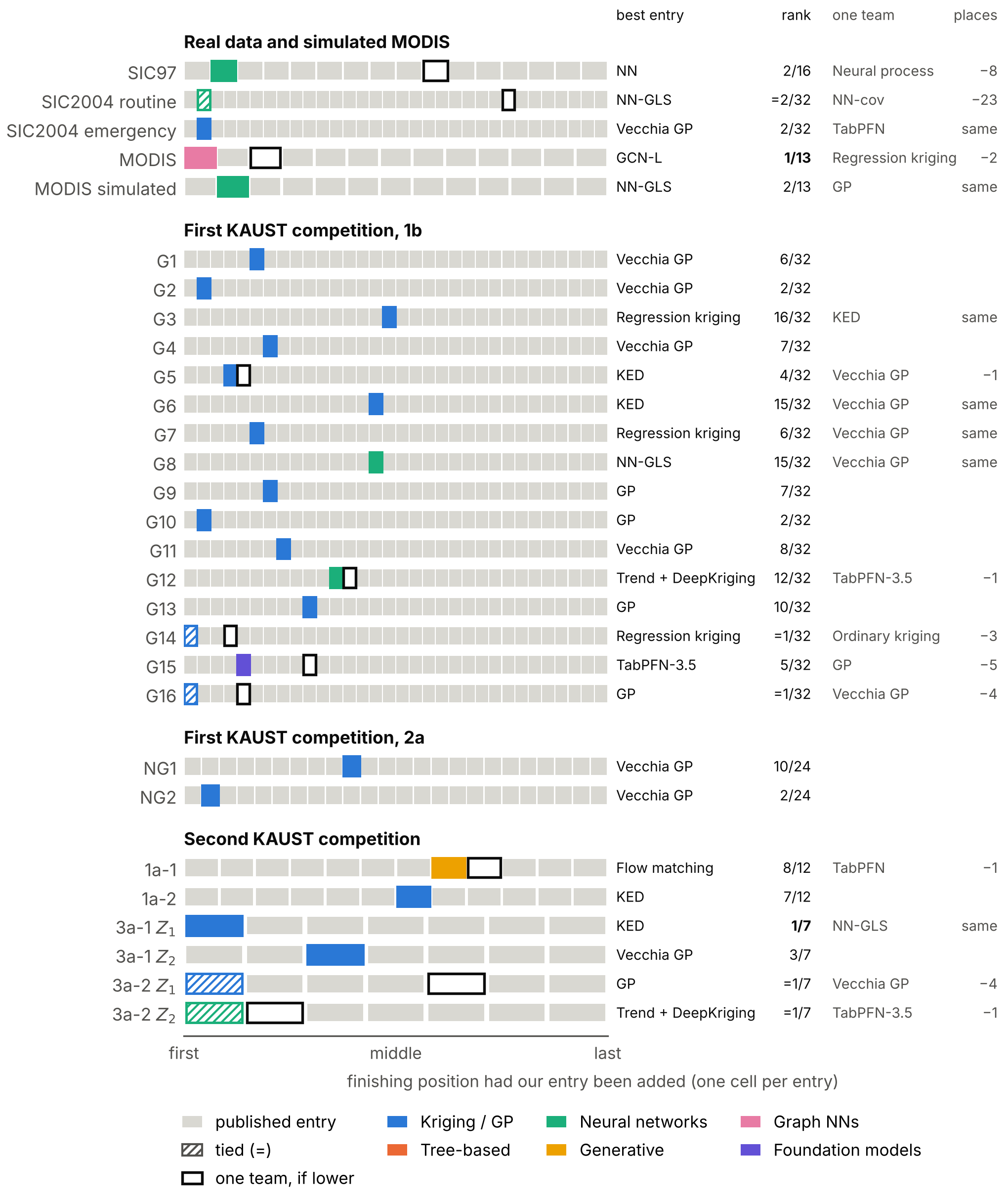}
\caption{Rank of our best 2026 entry on each leaderboard had it been added to the competition, with the configuration
selected on the training data: each row has one cell per entry (the published entries plus ours), from first to last; the
colored cell is our best entry, in the color of its family, and its name and rank $r/N$ are on the right. Hatched cells
and `=' mark a tie, at the published precision or, on G16 and 3a-2, by the true-model rule (main text, Section~3). One team: the single method we would have submitted
entering as one team, the tuned model with the lowest inner cross-validation score (Section~\ref{sec:tune}); where it
differs from our best entry it is named with the places it loses (``same'' if none), and an outlined cell marks where it
places; where it is our best entry, nothing is shown. In the places column, $-k$ means that the single method would have
placed $k$ places lower. A bold rank marks a first place that is not a tie. Scores are RMSE, except MAE for SIC2004. Where no
winner was named (SIC97, MODIS and its simulated version) and on each field of the first KAUST competition, the original
winner is the published entry with the lowest score.\label{fig:finishing}}
\end{figure}

\begin{figure}[!htbp]\centering
\includegraphics[width=\textwidth]{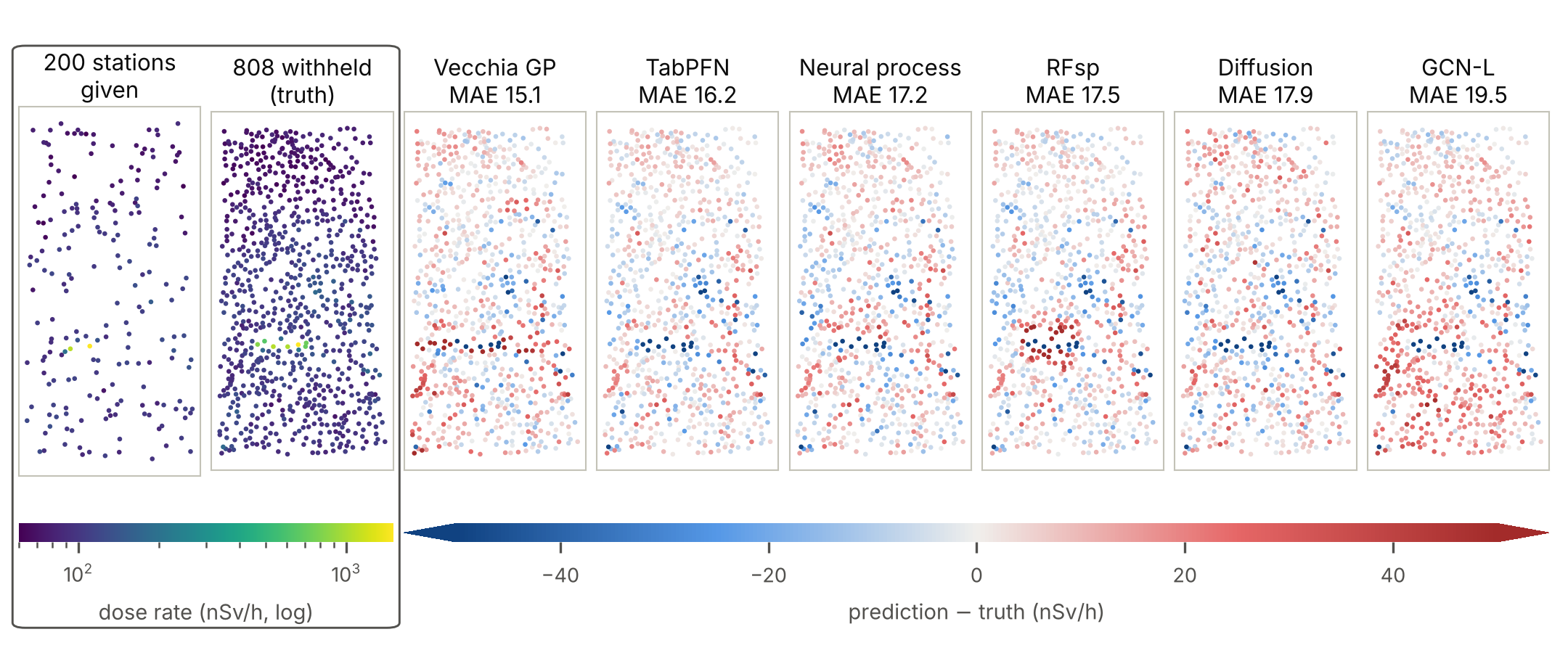}
\caption{SIC2004, emergency scenario: gamma dose rate at the 200 stations given to the participants and at the 808
withheld stations (logarithmic color scale), and the prediction error (prediction minus truth, nSv/h; errors beyond
$\pm50$ drawn at the ends of the scale) at the withheld stations of the best 2026 entry of each family
(main text, Section~3), ordered from best to worst test MAE.\label{fig:sic2004e}}
\end{figure}

\begin{figure}[!htbp]\centering
\includegraphics[width=\textwidth]{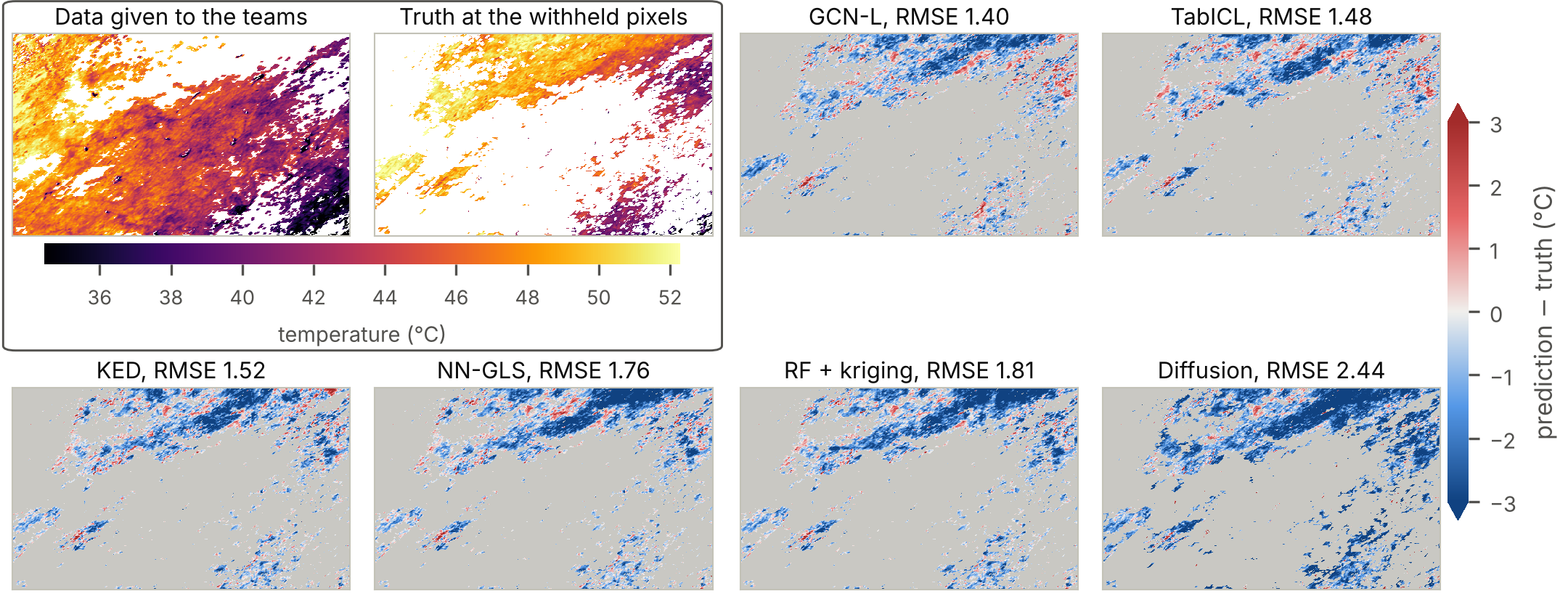}
\caption{MODIS land-surface temperature \citep{heaton2019}: the pixels released to the participants (white: withheld
under clouds), the truth at the withheld pixels, and the prediction error (prediction minus truth, $^\circ$C; errors beyond
$\pm3\,^\circ$C drawn at the ends of the scale) at the withheld pixels of the best 2026 entry of each family
(main text, Section~3), ordered from best to worst test RMSE; gray: pixels
released to the participants, which are not scored.\label{fig:modis}}
\end{figure}

Figures~\ref{fig:leaderboards_all} and \ref{fig:finishing} show all twenty-nine leaderboards, and
Figures~\ref{fig:sic2004e} and \ref{fig:modis} map the prediction errors in the SIC2004 emergency and on MODIS;
Tables~\ref{tab:kaust1} and \ref{tab:kaust} give one row per leaderboard of the two KAUST competitions, and
Table~\ref{tab:kaustb} the larger KAUST2 datasets. The repository
holds the score of every predictor (\texttt{results\_best/<dataset>/metrics.csv}) and the full ranking of every published
and 2026 entry on SIC97, SIC2004, MODIS and simulated MODIS (\texttt{paper/figs/rank\_<dataset>.png}). As
in the main text, each 2026 entry is the configuration selected by cross-validation on the training data, blind to the
test set (the default for the models that are not tuned); Section~\ref{sec:tuning} compares this choice with the default
and with the better of the two.

Predictor names are those of the main text (Table~2); GCN-L is the graph network with masked labels. Ranks compare our unrounded scores with the published ones, and ties are as defined in the main text (Section~3). A difference is
\emph{resolved} when its 95\% block-bootstrap interval excludes zero (Section~\ref{sec:boot}); the label describes the
sampling variability of our score against a fixed published number and is not a formal test. deepspat failed to fit on
eleven leaderboards (Section~\ref{sec:settings}) and has no score there. The published
MODIS entries (SPDE, NNGP, MRA, LAGP, FRK and the others) are described in \citet{heaton2019}.

\paragraph{SIC97.} The scores of the fifteen recovered entries are consistent with the summary of all submissions
reproduced by \citet{gilardi2003}. The lowest RMSE belongs to a multiquadric (radial basis function) interpolator
\citep[5.31\,mm;][]{thieken2003}. Our best entry, NN, would place second (5.36\,mm), and so would several others,
including KED and ordinary kriging. Elevation is an optional covariate: inner
cross-validation keeps it for seven of the twenty predictors that can use it and drops it for the others; for most
of them the choice changes the test score by less than 5\%. The only RMSE below the multiquadric's that we are aware of
is the post-competition mixture of four machine-learning models of \citet{gilardi2003}, 5.18\,mm. At the bottom of the
field, the ``simple good sense'' entry of \citet{genton2003}, in which Marc Genton estimated the 367 values by eye from the
100 observations and his knowledge of Swiss topography, did better than our random forest.

\paragraph{SIC2004.} On the routine day, twenty-seven of the thirty-one submissions were within 10\% of each other, and so
are all our entries except deepspat. NN-GLS has our best MAE and would place second, tied at the published precision with
the second-placed entry. On this day the variograms fitted for KED and regression kriging have little or no spatial
structure, and the Vecchia GP selects ranges far below the station spacing, so these entries predict essentially the
linear trend. In the emergency scenario the original winner, a general regression neural network \citep{timonin2005},
remains unbeaten in MAE, the organizers' ranking score. Our best entry, the Gaussian process with Vecchia approximation
(Vecchia GP), is second, 1.9\% behind: it maps part of the release, whereas TabPFN, which would also place second on its
own, shrinks toward
the background (Figure~\ref{fig:sic2004e}). This result is fragile. It rests on the seven withheld release stations, and its 95\% interval is wide
(Figure~\ref{fig:bootstrap}). The Gaussian process, which also selects an exponential covariance ($\nu=1/2$) but has a
different trend and estimation, scores far worse, so the result depends strongly on how the covariance is estimated. The other kriging and Gaussian process entries place between 14th and 23rd of 32.

\paragraph{MODIS.} Figure~\ref{fig:modis} maps the prediction errors of the best entry of each family. Three of our entries have a lower RMSE than the SPDE approach: GCN-L, TabICL with local context and KED.
The last two depend on choices made without test scores (Section~\ref{sec:tune}): the local
context of TabICL, chosen by inner cross-validation (Section~\ref{sec:settings}), and the cloud-mask folds of KED, chosen
by the rule that uses the locations only. Regression kriging, the method we would have
submitted entering as one team (Figure~\ref{fig:finishing}), would place third.

\paragraph{Simulated MODIS dataset.} On the simulated version of the MODIS data, a
stationary Gaussian field with constant mean on the same grid and with the same gaps, the 2019 multi-resolution approximation
entry has the lowest published RMSE. Five of our entries would place second: NN-GLS, the Gaussian process, the Vecchia GP, and ordinary and regression
kriging. The three leading entries on the real field fall back, TabICL to sixth, KED to seventh and GCN-L to tenth, the
same place as diffusion, the best generative model.

\paragraph{First KAUST competition.} The organizers' assessment sheet \citep{huang2021} lists thirty-one submissions in 1b and
twenty-three in 2a. On the
sixteen Gaussian fields of 1b the best submissions are within 0.11\% of the organizers' prediction with the true model and
parameters, and below it on fourteen of them, as expected of the best of thirty-one submissions scored on a single realization. The winner of 2a overall, which summed
the ranks on the two non-Gaussian fields, was a Tukey $g$-and-$h$ transformation followed by Gaussian process prediction
with Vecchia approximation; its bootstrap variant has the lowest RMSE on NG1, and a hierarchical-matrix Gaussian process
approximation (HMatrix) on NG2; RESSTE is the transformed Vecchia entry. Table~\ref{tab:kaust1} summarizes every field.

On the sixteen Gaussian fields of 1b our best entry is within 0.5\% of the original winner and of the true-model prediction
on all but G4, and it is a kriging or Gaussian process entry on thirteen of them. On G4, the smoothest field without a nugget, the Vecchia GP is 3.7\% behind the original winner, the only gap the bootstrap resolves
(Figure~\ref{fig:bootstrap}). The best entry outside kriging and Gaussian processes is close behind on the fields with a nugget or low smoothness and
far behind on the smooth fields without a nugget (G1, G2, G4, G5 and G7).

On the two non-Gaussian fields of 2a our best entry is the Vecchia GP, without transformation: tenth of 24 on NG1 and
second on NG2, behind only the hierarchical-matrix submission and ahead of every submission that transformed the data. The other families are behind on both fields, far behind on NG2. The paired bootstrap puts even the closest of them,
NN-GLS, behind the Vecchia GP. The heavy tails of these fields make the marginal intervals wide.

\paragraph{Second KAUST competition.} Its fields (Table~\ref{tab:kaust}) are Gaussian processes whose covariance class was announced, so the
teams entered purpose-built Gaussian process approximations. On three of the four 3a leaderboards our best entry is a kriging or Gaussian process entry: KED is first on 3a-1 $Z_1$, by
a margin the bootstrap resolves, and the Vecchia GP third on 3a-1 $Z_2$; on 3a-2 our best entries, the Gaussian process
and Trend + DeepKriging, tie the original winners (main text, Section~4.1). Sub-competition 1a is designed to defeat stationary methods: on
1a-1 our best entry, flow matching, is eighth, while the teams that modeled the varying mean lead; on 1a-2 KED is seventh.
TabPFN-3.5 is usually the strongest foundation model, but it is never our best entry on these leaderboards.
Flexible methods depend strongly on
their configuration: on 3a-1 $Z_2$,
DeepKriging-P with the authors' default settings has an RMSE more than seven times the original winner's, and only tuning
on the training data brings it to 33\% above the winner's RMSE.

\paragraph{KAUST2 1b and 3b (larger datasets).} These sub-competitions repeat
KAUST2 1a and 3a with ten times as many training locations (Table~\ref{tab:kaustb}); they are distinct from the sixteen
KAUST1 1b fields. We entered every predictor at its default
configuration, without tuning, so these entries are not comparable with the tuned entries of the twenty-nine leaderboards;
they check whether the picture changes with the size of the data. It does not: on seven of the eight leaderboards both our
best entry and our best kriging or Gaussian process entry are within 1\% of the original winner. TabPFN-3.5 scores
slightly below the original winner on the second variable of 3b-2 and of 3b-3, by margins the bootstrap does not resolve.
The original winner is ahead of every 2026 entry by a resolved margin only on 3b-1: on its first variable by less than
1\%, and on the second, a field with very little noise, the two leading teams are about 20\% ahead of every 2026 entry,
and our best entry places third.
The fields of 3b-2 and 3b-3 are not independent: listed in the order of the solution file, their test values correlate
row by row (0.91 and 0.96 for the two variables), although the locations in the two files differ, which suggests a
shared noise seed.

\begin{table}[!t]\centering\footnotesize\setlength{\tabcolsep}{1.8pt}
\caption{First KAUST competition \citep{huang2021}: the sixteen Gaussian fields of 1b (31 submissions) and the two non-Gaussian fields of 2a (23 submissions), 90{,}000 training and 10{,}000 test locations each. Columns, relative to the original winner's RMSE (the lowest-RMSE submission on each field): the organizers' prediction with the true model and parameters (1b), our best entry, our best kriging or Gaussian process entry and our best entry of any other family (baselines excluded), with the rank among the submissions (`=' marks a tie, main text Section~3). Field parameters of 1b (organizers' design; variance 1.5): Mat\'ern smoothness $\nu$, effective range (distance at which the correlation falls to 5\%) and nugget variance. Team names are shortened to the part before any parenthesis. The score of every predictor is in \texttt{results\_best/<key>/metrics.csv} in the repository.}
\label{tab:kaust1}
\begin{tabular}{lrrr>{\raggedright\arraybackslash}p{2.3cm}rlrrrr}\toprule
 & \multicolumn{3}{c}{Field} & Original & True & \multicolumn{3}{c}{Our best entry} & Best kriging/GP & Best other \\ \cmidrule(lr){2-4}\cmidrule(lr){7-9}
Leaderboard & $\nu$ & Range & Nugget & winner & model & Model & vs winner & Rank & vs winner & vs winner \\ \midrule
G1 & 2.3 & 0.1 & 0 & RESSTE & +0.02\% & Vecchia GP & +0.44\% & 6/32 & +0.44\% & +104.9\% \\
G2 & 1.5 & 0.1 & 0 & UNL-\allowbreak{}Stats & +0.02\% & Vecchia GP & +0.02\% & 2/32 & +0.02\% & +21.9\% \\
G3 & 0.6 & 0.1 & 0 & Team-\allowbreak{}MS & +0.006\% & Regression kriging & +0.25\% & 16/32 & +0.25\% & +0.26\% \\
G4 & 2.3 & 0.3 & 0 & HCHISS & +0.11\% & Vecchia GP & +3.7\% & 7/32 & +3.7\% & +253.9\% \\
G5 & 1.5 & 0.3 & 0 & ExtStat & +0.02\% & KED & +0.02\% & 4/32 & +0.02\% & +35.5\% \\
G6 & 0.6 & 0.3 & 0 & AppStatUZH & +0.03\% & KED & +0.16\% & 15/32 & +0.16\% & +0.18\% \\
G7 & 1.5 & 0.8 & 0 & GpGp & +0.02\% & Regression kriging & +0.05\% & 6/32 & +0.05\% & +36.5\% \\
G8 & 0.6 & 0.8 & 0 & RESSTE & +0.01\% & NN-GLS & +0.15\% & 15/32 & +0.15\% & +0.15\% \\
G9 & 2.3 & 0.1 & 0.27 & RESSTE & $-$0.0001\% & GP & +0.02\% & 7/32 & +0.02\% & +0.47\% \\
G10 & 1.5 & 0.1 & 0.27 & Team-\allowbreak{}MS & +0.004\% & GP & +0.004\% & 2/32 & +0.004\% & +0.44\% \\
G11 & 0.6 & 0.1 & 0.27 & AppStatUZH & +0.001\% & Vecchia GP & +0.03\% & 8/32 & +0.03\% & +0.13\% \\
G12 & 2.3 & 0.3 & 0.27 & SpatStat-\allowbreak{}Fans & +0.01\% & Trend + DeepKriging & +0.11\% & 12/32 & +0.17\% & +0.11\% \\
G13 & 1.5 & 0.3 & 0.27 & RESSTE & +0.006\% & GP & +0.10\% & 10/32 & +0.10\% & +0.19\% \\
G14 & 0.6 & 0.3 & 0.27 & HCHISS & $-$0.005\% & Regression kriging & $-$0.000005\% & =1/32 & $-$0.000005\% & +0.08\% \\
G15 & 1.5 & 0.8 & 0.27 & ExtStat & +0.02\% & TabPFN-3.5 & +0.03\% & 5/32 & +0.05\% & +0.03\% \\
G16 & 0.6 & 0.8 & 0.27 & Among-\allowbreak{}Stats & +0.04\% & GP & $-$0.003\% & =1/32 & $-$0.003\% & +0.08\% \\
NG1 & -- & -- & -- & RESSTE & -- & Vecchia GP & +2.7\% & 10/24 & +2.7\% & +6.7\% \\
NG2 & -- & -- & -- & HMatrix & -- & Vecchia GP & +0.65\% & 2/24 & +0.65\% & +23.1\% \\
\bottomrule\end{tabular}\end{table}

\begin{table}[!t]\centering\footnotesize\setlength{\tabcolsep}{3pt}
\caption{Second KAUST competition \citep{abdulah2022}, sub-competitions 1a (11 teams) and 3a (6 teams): 90{,}000 training and 10{,}000 test locations in 1a, 45{,}000 and 5{,}000 in 3a, where the other variable of the pair is the optional covariate. Columns, relative to the original winner's RMSE: the organizers' prediction with the true model and parameters (where published), our best entry, our best kriging or Gaussian process entry and our best entry of any other family (baselines excluded), with the rank among the published teams (`=' marks a tie, main text Section~3). Team names are shortened to the part before any parenthesis. The score of every predictor is in \texttt{results\_best/<key>/metrics.csv} in the repository.}
\label{tab:kaust}
\begin{tabular}{l>{\raggedright\arraybackslash}p{3.4cm}rlrrrr}\toprule
 & Original & True & \multicolumn{3}{c}{Our best entry} & Best kriging/GP & Best other \\ \cmidrule(lr){4-6}
Leaderboard & winner & model & Model & vs winner & Rank & vs winner & vs winner \\ \midrule
1a-1 & Spatial Special & -- & Flow matching & +1.0\% & 8/12 & +1.3\% & +1.0\% \\
1a-2 & GeoModels & -- & KED & +1.1\% & 7/12 & +1.1\% & +1.2\% \\
3a-1 $Z_1$ & GpGp & $-$5.9\% & KED & $-$2.0\% & 1/7 & $-$2.0\% & $-$1.9\% \\
3a-1 $Z_2$ & GpGp & $-$5.3\% & Vecchia GP & +0.17\% & 3/7 & +0.17\% & +18.6\% \\
3a-2 $Z_1$ & GeoModels & +0.01\% & GP & $-$0.0006\% & =1/7 & $-$0.0006\% & +0.04\% \\
3a-2 $Z_2$ & Spatial Special & +0.03\% & Trend + DeepKriging & $-$0.03\% & =1/7 & +0.20\% & $-$0.03\% \\
\bottomrule\end{tabular}\end{table}

\begin{table}[!t]\centering\scriptsize\setlength{\tabcolsep}{3pt}
\caption{Sub-competitions 1b and 3b of the second KAUST competition \citep{abdulah2022}: ten times the training size of 1a and 3a (900{,}000 training locations in 1b, 450{,}000 in 3b), with every predictor at its default configuration (not tuned); -cov: with the other variable as the optional covariate. RMSE of our best entry and of our best kriging or Gaussian process entry relative to the original winner (\%), with the marginal 95\% block-bootstrap interval of our best entry as in Figure~\ref{fig:bootstrap}, and the rank among the published teams ($N$: teams plus one). deepspat failed on all eight. In 3b every predictor that accepts a covariate uses the other variable of the pair (without tuning there is no choice with or without it).}
\label{tab:kaustb}
\begin{tabular}{lllrrlrr}\toprule
 & Original & \multicolumn{3}{c}{Our best entry} & \multicolumn{3}{c}{Best kriging/GP} \\ \cmidrule(lr){3-5}\cmidrule(lr){6-8}
Leaderboard & winner & Model & vs winner (\%) & Rank & Model & vs winner (\%) & Rank \\ \midrule
1b-1 & Spatial Special & TabPFN-3.5 & +0.23 [$-$0.3, +0.7] & 3/11 & GP & +0.83 & 9/11 \\
1b-2 & Spatial Special & KED & +0.97 [$-$14.4, +16.4] & 6/11 & KED & +0.97 & 6/11 \\
3b-1 $Z_1$ & JPTIKKS & NN-GLS & +0.82 [+0.2, +1.4] & 2/7 & KED & +0.85 & 2/7 \\
3b-1 $Z_2$ & JPTIKKS & NN-GLS & +20.7 [+19.8, +21.7] & 3/7 & Ordinary kriging & +21.6 & 3/7 \\
3b-2 $Z_1$ & AppStatUZH & GP & +0.03 [$-$0.5, +0.6] & 2/7 & GP & +0.03 & 2/7 \\
3b-2 $Z_2$ & Spatial Special & TabPFN-3.5 & $-$0.02 [$-$0.5, +0.5] & 1/7 & Vecchia GP & +0.25 & 4/7 \\
3b-3 $Z_1$ & Spatial Special & DeepKriging-cov & +0.07 [$-$0.4, +0.6] & 3/7 & GP & +0.29 & 4/7 \\
3b-3 $Z_2$ & Envstat.ai & TabPFN-3.5 & $-$0.17 [$-$0.8, +0.5] & 1/7 & GP & +0.22 & 4/7 \\
\bottomrule\end{tabular}\end{table}

\clearpage

\FloatBarrier
\section{Sampling uncertainty, tuning and selection}\label{sec:tuning}
\subsection{Uncertainty in the scores}\label{sec:boot}
Each leaderboard is scored on a single test set, and a different but similar test set would give somewhat different
scores. We measure this with a spatial block bootstrap of the test locations \citep{lahiri2003}. The test locations are
grouped once into fifty spatially compact blocks by $k$-means clustering of their coordinates. Each bootstrap sample draws fifty of these blocks with replacement and pools their locations;
whole blocks are drawn, rather than single locations, because errors at nearby locations are correlated: resampling
single locations would treat them as independent and give intervals that are too narrow. We draw 10{,}000 bootstrap
samples and report percentile intervals, the 2.5th and 97.5th percentiles of the resampled ratios. The Monte Carlo
standard error of an interval limit, from ten batches of 1{,}000 resamples, is at most 1.2 percentage points, but much
smaller for the limits that decide whether a difference is resolved: 0.03 for the upper limit of KED on KAUST2 3a-1
$Z_1$, 0.05 for the lower limit of the Vecchia GP on KAUST1 G4, and 0.12 and 0.08 for the upper limits of GCN-L against
the SPDE approach and against KED on MODIS. The
intervals keep each fitted predictor fixed; they do not include the variability from refitting it.

To compare one of our predictors with the winner, whose predictions are not public, we recompute our score on each
resample and divide it by the winner's published score: the central 95\% of these ratios is the \emph{marginal}
interval. It describes the sampling variability of our score against a fixed published number; the winner's own
variability cannot be included. To compare two of our predictors, we score both on the same resample and divide one
score by the other: the \emph{paired} interval. Paired intervals are much narrower, because most of the variation cancels
in the ratio, so a paired comparison with the winners \citep{diebold1995,hering2011}, which would need their predictions,
would probably be sharper, and a difference that our marginal interval does not resolve might still be real. As in the
main text, a difference is \emph{resolved} if the 95\% interval of the ratio excludes one; the figures and tables give
the ratios as percentage differences, whose intervals then exclude zero.

\begin{figure}[p]\centering
\includegraphics[width=\textwidth,height=0.6\textheight,keepaspectratio]{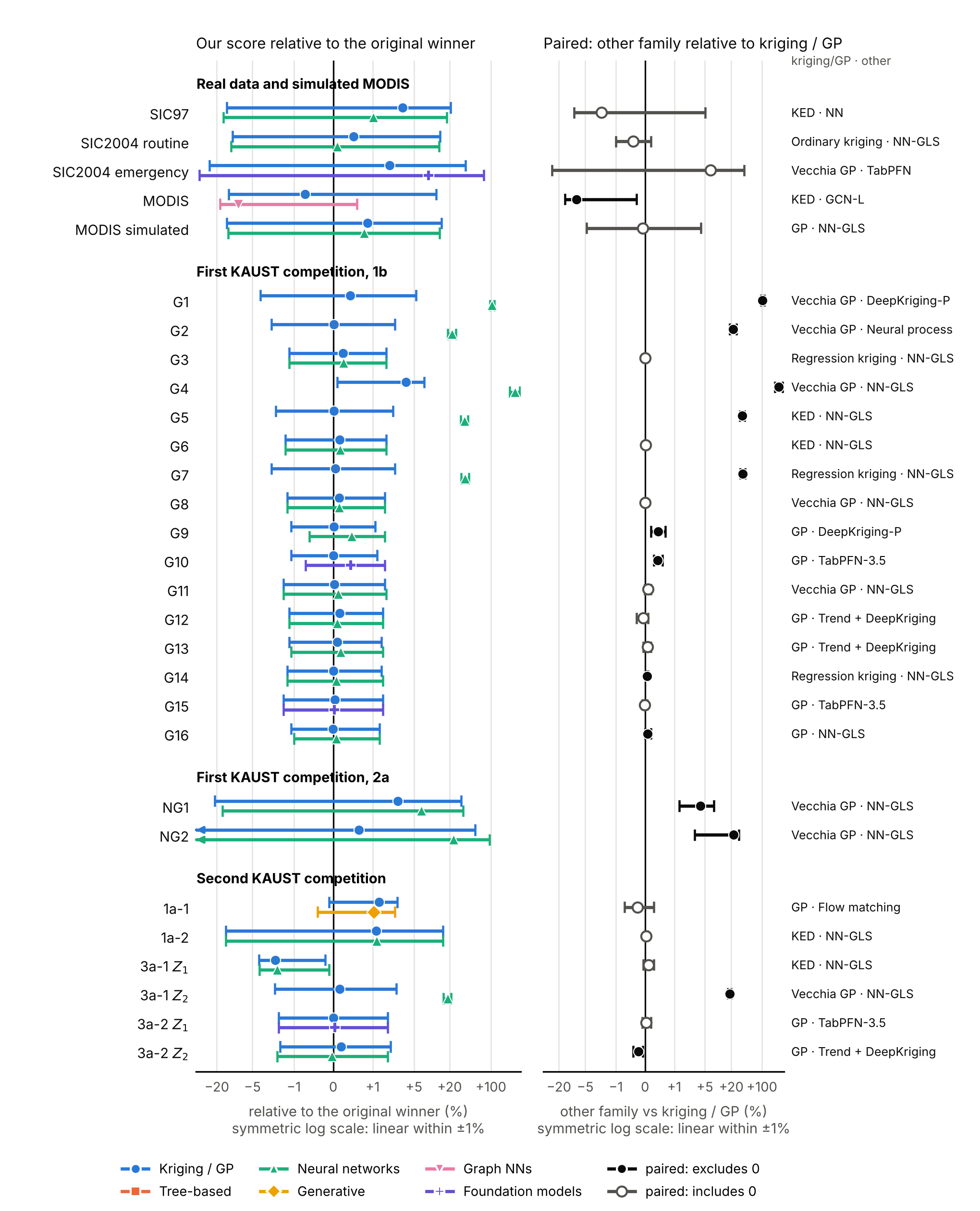}
\caption[Sampling uncertainty of the 2026 scores]{\input{tables/bootstrap_caption}\label{fig:bootstrap}}
\end{figure}

\paragraph{Intervals.} Figure~\ref{fig:bootstrap} gives the marginal intervals of the best kriging or Gaussian process
entry and of the best entry of any other family, and the paired interval between the two. On SIC97, SIC2004 and MODIS
the marginal intervals are wide, so a single test set has little power to separate the leading entries from the original
winners. On the second KAUST competition's fields they are narrow, except on 1a-2, whose covariance varies across the
domain by design; on the Gaussian fields of the first they are a few percent wide, and on its non-Gaussian 2a fields much
wider, because of their heavy tails. Paired differences between our entries are usually
much narrower; the SIC2004 emergency, dominated by seven stations, is the exception.

\paragraph{Number of blocks.} Fifty blocks hold about 7 test stations each on SIC97, 16 on SIC2004, 100 to 200 on the
KAUST fields and about 850 on MODIS. Resampled blocks are close to independent only if they are wider than the distance
over which the errors are correlated. For our best entry on each leaderboard, Table~\ref{tab:bootblocks_all} estimates
that distance from the empirical correlogram of its errors (all pairs within a random subsample of at most 3{,}000 test
locations; the mean product of the centered errors in 40 distance bins of equal width up to twice the block width,
divided by their variance) as the shortest distance at which the
correlation falls below 0.1, and compares it with the median width of the 50 blocks (the width of a block is twice the
root-mean-square distance of its locations from its centroid). On the KAUST fields the correlation
falls below 0.1 within the shortest distances resolved, about the spacing of the test locations, so the blocks are at
least 10 times wider than the correlation distance; on SIC2004 they are 3 to 4 times wider, but on SIC97, MODIS and its
simulated version only 1.0 to 1.5 times. There neighboring blocks are correlated, the intervals may be somewhat too
narrow, and more, smaller blocks make them less trustworthy, not more.

The table repeats the marginal interval of our best entry with 25, 50 and 100 blocks, with 100{,}000 resamples each, so
that Monte Carlo error (at most 0.36 percentage points) no longer matters. Which differences are resolved depends on the
number of blocks on two leaderboards. On KAUST2 3a-1 $Z_1$ the interval of KED excludes zero with 50 blocks, only just
with 100 (upper limit $-0.01\%$) and not with 25 ($+0.01\%$). On MODIS that of GCN-L excludes zero only with 100 blocks,
the setting least to be trusted there, since its blocks are narrower than the correlation distance; with 25 or 50 blocks
it does not. On KAUST1 G4 the interval excludes zero with every number of blocks, and on twenty-six leaderboards with
none. On SIC97 and routine SIC2004 the intervals narrow somewhat as the blocks get smaller, but remain far from excluding
zero; in the SIC2004 emergency the seven withheld release stations fall in three of the fifty blocks (two of 25, four of 100), and the interval stays wide
with any number of blocks. On MODIS, every interval for KED against the SPDE approach covers zero with 10 to
100 blocks, while the paired interval of GCN-L against KED excludes zero with 10, 50 and 100 blocks but not with 25
(\texttt{paper/ADDITIONAL\_RESULTS.md}). We therefore call both comparisons involving GCN-L borderline. Resampling the
cloud gaps themselves is not usable: one of the 563 gaps holds most of the test pixels.

\begin{table}[!t]\centering\footnotesize\setlength{\tabcolsep}{3.5pt}
\caption{Sensitivity of the block bootstrap to the number of $k$-means blocks, on every leaderboard: marginal 95\% percentile interval (\%) of the score of our best entry relative to the original winner's published score, with 25, 50 and 100 blocks of the test locations (100{,}000 resamples each, drawn independently of the 10{,}000 of Figure~\ref{fig:bootstrap}, so the 50-block column differs from it by Monte Carlo error; the Monte Carlo standard error of every limit here is at most 0.36 percentage points; limits within one percentage point of zero are given to two decimals). $n$: test locations; vs winner: the score of our best entry relative to the original winner (\%); width/corr.: median width of the 50 blocks divided by the distance at which the correlation of that entry's errors falls below 0.1 ($\geq$: the correlation falls below 0.1 within the shortest distances resolved, about the spacing of the test locations, so the ratio is a lower bound). Bold: the interval excludes zero.}\label{tab:bootblocks_all}
\begin{tabular}{llrrrlll}\toprule
Leaderboard & Best entry & $n$ & vs winner & width/corr. & 25 blocks & 50 blocks & 100 blocks \\ \midrule
SIC97 & NN & 367 & $+1.0$ & 1.2 & $[-16.9, +17.8]$ & $[-15.3, +18.2]$ & $[-13.4, +15.5]$ \\
SIC2004 routine & NN-GLS & 808 & $+0.10$ & 3.3 & $[-13.1, +14.8]$ & $[-11.5, +13.3]$ & $[-9.6, +10.7]$ \\
SIC2004 emergency & Vecchia GP & 808 & $+1.9$ & 4.0 & $[-34.3, +49.3]$ & $[-26.1, +37.7]$ & $[-25.2, +38.3]$ \\
MODIS & GCN-L & 42{,}740 & $-8.7$ & 1.0 & $[-19.7, +1.7]$ & $[-17.9, +0.52]$ & {\boldmath$[-15.7, -1.6]$} \\
MODIS simulated & NN-GLS & 44{,}431 & $+0.78$ & 1.5 & $[-13.9, +15.3]$ & $[-12.5, +13.5]$ & $[-10.2, +12.1]$ \\
KAUST1 1b G1 & Vecchia GP & 10{,}000 & $+0.44$ & $\geq$10.0 & $[-3.4, +5.1]$ & $[-3.8, +5.4]$ & $[-3.5, +5.1]$ \\
KAUST1 1b G2 & Vecchia GP & 10{,}000 & $+0.02$ & $\geq$10.0 & $[-2.5, +2.4]$ & $[-2.5, +2.4]$ & $[-2.4, +2.5]$ \\
KAUST1 1b G3 & Regression kriging & 10{,}000 & $+0.25$ & $\geq$10.0 & $[-1.6, +2.0]$ & $[-1.2, +1.7]$ & $[-1.4, +2.0]$ \\
KAUST1 1b G4 & Vecchia GP & 10{,}000 & $+3.7$ & $\geq$10.0 & {\boldmath$[+0.43, +7.2]$} & {\boldmath$[+0.12, +7.5]$} & {\boldmath$[+0.18, +7.4]$} \\
KAUST1 1b G5 & KED & 10{,}000 & $+0.02$ & $\geq$10.0 & $[-1.9, +1.9]$ & $[-2.0, +2.2]$ & $[-2.1, +2.2]$ \\
KAUST1 1b G6 & KED & 10{,}000 & $+0.16$ & $\geq$10.0 & $[-1.6, +1.8]$ & $[-1.4, +1.7]$ & $[-1.4, +1.7]$ \\
KAUST1 1b G7 & Regression kriging & 10{,}000 & $+0.05$ & $\geq$10.0 & $[-2.6, +2.9]$ & $[-2.3, +2.3]$ & $[-2.3, +2.4]$ \\
KAUST1 1b G8 & NN-GLS & 10{,}000 & $+0.15$ & $\geq$10.0 & $[-1.3, +1.6]$ & $[-1.3, +1.6]$ & $[-1.3, +1.7]$ \\
KAUST1 1b G9 & GP & 10{,}000 & $+0.02$ & $\geq$20.0 & $[-1.2, +1.2]$ & $[-1.1, +1.1]$ & $[-1.3, +1.4]$ \\
KAUST1 1b G10 & GP & 10{,}000 & $+0.00$ & $\geq$20.0 & $[-1.2, +1.2]$ & $[-1.1, +1.2]$ & $[-1.3, +1.3]$ \\
KAUST1 1b G11 & Vecchia GP & 10{,}000 & $+0.03$ & $\geq$20.0 & $[-1.6, +1.6]$ & $[-1.6, +1.6]$ & $[-1.5, +1.5]$ \\
KAUST1 1b G12 & Trend + DeepKriging & 10{,}000 & $+0.11$ & $\geq$20.0 & $[-1.2, +1.5]$ & $[-1.2, +1.5]$ & $[-1.3, +1.5]$ \\
KAUST1 1b G13 & GP & 10{,}000 & $+0.10$ & $\geq$20.0 & $[-0.97, +1.2]$ & $[-1.2, +1.4]$ & $[-1.1, +1.3]$ \\
KAUST1 1b G14 & Regression kriging & 10{,}000 & $-0.00$ & $\geq$20.0 & $[-1.2, +1.2]$ & $[-1.3, +1.4]$ & $[-1.3, +1.3]$ \\
KAUST1 1b G15 & TabPFN-3.5 & 10{,}000 & $+0.03$ & $\geq$20.0 & $[-1.5, +1.6]$ & $[-1.5, +1.5]$ & $[-1.5, +1.5]$ \\
KAUST1 1b G16 & GP & 10{,}000 & $-0.00$ & $\geq$10.0 & $[-1.0, +1.0]$ & $[-1.1, +1.2]$ & $[-1.4, +1.4]$ \\
KAUST1 2a NG1 & Vecchia GP & 10{,}000 & $+2.7$ & $\geq$10.0 & $[-23.6, +31.3]$ & $[-21.7, +31.5]$ & $[-18.0, +24.8]$ \\
KAUST1 2a NG2 & Vecchia GP & 10{,}000 & $+0.65$ & $\geq$10.0 & $[-58.0, +57.3]$ & $[-53.3, +54.2]$ & $[-41.7, +38.2]$ \\
KAUST2 1a-1 & Flow matching & 10{,}000 & $+1.0$ & $\geq$20.0 & $[-0.27, +2.2]$ & $[-0.39, +2.4]$ & $[-0.36, +2.4]$ \\
KAUST2 1a-2 & KED & 10{,}000 & $+1.1$ & $\geq$10.0 & $[-21.0, +21.8]$ & $[-14.0, +15.7]$ & $[-9.5, +11.5]$ \\
KAUST2 3a-1 $Z_1$ & KED & 5{,}000 & $-2.0$ & $\geq$10.0 & $[-4.0, +0.01]$ & {\boldmath$[-3.9, -0.24]$} & {\boldmath$[-4.0, -0.01]$} \\
KAUST2 3a-1 $Z_2$ & Vecchia GP & 5{,}000 & $+0.17$ & $\geq$10.0 & $[-2.4, +2.8]$ & $[-2.1, +2.5]$ & $[-2.3, +2.6]$ \\
KAUST2 3a-2 $Z_1$ & GP & 5{,}000 & $-0.00$ & $\geq$20.0 & $[-2.1, +2.0]$ & $[-1.8, +1.8]$ & $[-1.8, +1.8]$ \\
KAUST2 3a-2 $Z_2$ & Trend + DeepKriging & 5{,}000 & $-0.03$ & $\geq$20.0 & $[-1.8, +1.7]$ & $[-1.9, +1.8]$ & $[-1.8, +1.8]$ \\
\bottomrule\end{tabular}\end{table}

\subsection{Tuning and selection}\label{sec:tune}
\paragraph{Protocol.} Every tuned predictor is scored on twenty configurations, and the tabular foundation models on thirteen
(the default and the twelve points of their search space, which include it; see Tuning coverage below). Each variant receives the full budget, so models with more variants are searched more: where a model has an optional covariate,
a choice of context (the foundation models on the large datasets, local or random; Section~\ref{sec:impl}) or a
learning-rate scale (deepspat), each variant is tuned with this budget, and the variant with the lowest inner
cross-validation score enters.
The twenty are the default and nineteen random draws from a model-specific search space (listed in the repository,
\texttt{paper/ADDITIONAL\_RESULTS.md}). Several of these search
spaces are effectively smaller than they appear. On SIC97 and SIC2004, with fewer than 1{,}000 training points, kriging
and the Gaussian process use all points, so the neighborhood size of kriging and of inverse distance weighting and the
subsample size and prediction neighbors of the Gaussian process have no effect there (the Gaussian process is then
tuned over its smoothness only, three distinct configurations).

On the large datasets (MODIS, its simulated version and the KAUST fields), several tuned settings are not used. The
generative models are trained without dropout and with weight decay $10^{-4}$, so their tuned dropout and weight decay
have no effect, and the coordinate networks are trained in mini-batches with a separately tuned learning rate, so their
full-batch learning rate has none. The empirical variogram is computed with fixed lags and cutoff (Section~\ref{sec:settings}), so the tuned number of
lags and cutoff are not used either. The number of ensemble members
of TabICL has almost no effect on its inner cross-validation scores (configurations that differ only in it agree to
within 0.025\% in relative terms on MODIS and the second KAUST competition), so there
TabICL is tuned effectively over the context size. On these datasets kriging evaluates between 9 and 15
distinct configurations of its twenty, because only the variogram family and the neighborhood size matter there. Each configuration is
scored by five-fold cross-validation on the training data only, with the competition metric, and the selected
configuration is refitted on all training data. On MODIS, where the test set is a set of cloud gaps, the cross-validation folds
imitate them: each of the five held-out sets is drawn independently, as the training pixels under a copy of the full
missing-pixel mask of the released image, shifted circularly at random and flipped at random along each axis
(19--34\% of the training pixels; copies holding out less than 10\% or more than 50\% are redrawn). The held-out sets
can overlap and need not cover every pixel, so on MODIS the cross-validation ``folds'' are five repeated hold-out sets rather than a
partition. The design is chosen by the rule of the main text (Section~3), which uses the locations only
\citep[see][for designs that match the prediction distances]{mila2022}: for each candidate design (random five-fold
cross-validation; five spatial blocks from $k$-means on the training coordinates; on MODIS and its simulated version, the
cloud-mask hold-outs), it computes the distance from each held-out point to the nearest point used for fitting, and
selects the design whose distribution of these distances is closest, in Wasserstein-1 distance divided by the mean
test distance, to the distribution of the distances from the test locations to the nearest training location. On MODIS
these relative distances are 0.84 (random), 8.0 ($k$-means blocks) and 0.14 (cloud mask), and on its simulated version
the cloud-mask hold-outs are again closest (0.12). On SIC97, SIC2004 and the KAUST2 fields random folds are closest
(0.07 to 0.13, against 3.2 to 67 for $k$-means blocks), and on the KAUST1 fields too (0.09 to 0.12, against 79 to 86;
\texttt{tools/fold\_match.py}).

\begin{figure}[p]\centering
\includegraphics[width=\textwidth,height=0.76\textheight,keepaspectratio]{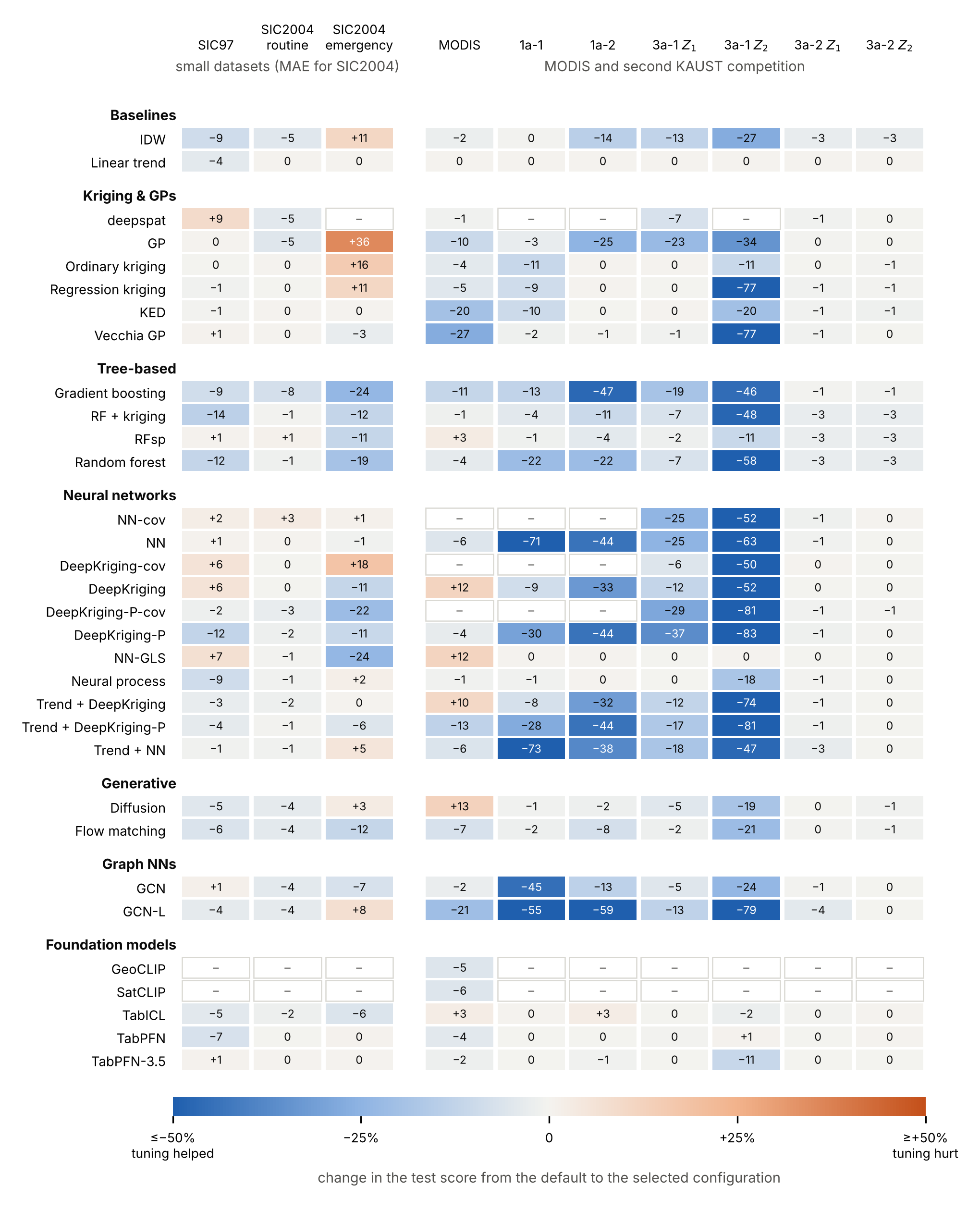}
\caption{Change in test score from tuning: relative change of the test score from the default configuration
to the configuration selected by tuning, for each predictor (rows, by family) and leaderboard (columns); blue: tuning
helped, orange: it hurt (colors saturate at $\pm50\%$); --: not run on that leaderboard or failed. Each predictor scores 20
configurations (13 for the tabular foundation models, of which only 4 are distinct on the small datasets, where every
context size exceeds the training set, and 12 on MODIS and the second KAUST competition) by five-fold cross-validation on
the training data only, with folds built from the real cloud masks on MODIS. Scores are RMSE, except MAE for SIC2004. The
linear trend is not tuned; on SIC97 its change comes from the choice of elevation as a covariate.\label{fig:tuned}}
\end{figure}

\paragraph{Results.} Figure~\ref{fig:tuned} compares the tuned and default entries on the ten
leaderboards of SIC97, SIC2004, MODIS and the second KAUST competition. The same comparison on the eighteen fields of the first KAUST
competition and on the simulated MODIS dataset, left out of the figure for legibility, is in the repository (the
summaries in \texttt{results2} and \texttt{results\_tuned}) and is counted in the overall numbers of the main text
(Section~4). Tuning helps most where the defaults are poorly
matched to the data: KED on MODIS moves from eighth to first among the published entries, and DeepKriging-P improves by 30\% or more on 1a and 3a-1, almost sixfold on
3a-1 $Z_2$. Across the 768 combinations of tuned predictor and leaderboard it lowers the test score in 85\% of cases (median change
over all 768
$-2.8\%$), and where it raises it the median loss is 0.5\%. It can also make an entry worse, as discussed below. How the cross-validation folds are built matters: tuned on
$k$-means blocks, KED would select a spherical variogram and score worse on MODIS (RMSE 1.556\,$^\circ$C, against
1.53\,$^\circ$C for the SPDE approach, which would place it second).

\paragraph{How well inner cross-validation ranks models.} Across the tuned models, the rank correlation between the inner
cross-validation score of the selected configuration and the test score (Spearman, over the 25 to 29 tuned predictors of each
leaderboard) is close to one on the second KAUST competition's fields (0.92 to 0.99), high on MODIS with the cloud-mask
folds (0.73 over its 28 tuned predictors), weak on SIC97 and routine SIC2004 (0.07 and 0.27) and
moderate in the SIC2004 emergency (0.53). An earlier set of MODIS runs tuned on $k$-means
blocks gave a negative correlation ($-0.35$ over 19 predictors), and with $k$-means blocks the test RMSE of those 19
tuned predictors differs from that with the cloud-mask folds by $-18\%$ to $+25\%$ (higher for 11); the two sets of
runs differ in other respects, so this comparison is indicative only (\texttt{results\_archive/}). Cross-validation folds that
imitate the task do not make the inner scores reliable estimates of the test error: on MODIS the inner errors are a median
21\% below the test errors.

\paragraph{When tuning chooses a worse configuration.} Selecting configurations on the training data is the only
legitimate choice in a competition, but it can pick a configuration that is worse on the test set. On MODIS, even with
folds built from the cloud masks, six models are worse after tuning, among them TabICL and NN-GLS. In the SIC2004 emergency
the release is absent from almost all training stations, so the selected configurations help some models and hurt others
markedly; ordinary kriging, for example, selects a configuration that predicts a constant. Such losses are part of the
primary, blind analysis; the better of the default and the tuned entry, which uses the test set, is a sensitivity
analysis only (below).

\paragraph{Tuning coverage.} On SIC97 and SIC2004, where every context size exceeds the training set, only the four
values of the number of ensemble members of the tabular foundation models differ. Every other predictor except the linear trend, deepspat included (Section~\ref{sec:impl}), was tuned on every leaderboard on which it was run, with
the same budget and rules. The coarser DeepKriging, and the graph networks on MODIS, its simulated version and the KAUST
fields, were tuned after the other results were known. A configuration still running after 20 hours of wall-clock time (on 16 CPU cores, or on one GPU) was
recorded as not completed and left out of the selection, and the final refit used the best completed configuration. The
cap was introduced during the study; applied to the earlier DeepKriging-P runs, it changes no selection. In the tuning runs that entered, the cap
stopped configurations of DeepKriging-P and Trend + DeepKriging-P (87), of deepspat (58 in the variants that entered,
72 of the 460 configurations of all its tuning runs, with and without the covariate; Section~\ref{sec:settings}) and of TabPFN-3.5 (one, on G3);
before the cap existed, one DeepKriging-P configuration on MODIS was stopped by hand, after 49 hours, and recorded in
the same way. For DeepKriging-P
the cap mostly removed its costliest configurations (400 epochs or the finest basis), so on the KAUST fields it is selected
among its cheaper configurations, which may disadvantage it. On MODIS, five of the twenty
configurations of GeoCLIP first failed on a corrupted read of a shared file; rerun, they leave its selection unchanged. 

\paragraph{Sensitivity to how entries are chosen.} The main text represents every model by the configuration selected on
the training data, blind to the test set. With every
model at its default, also blind, the placement of our best entry is the same on thirteen leaderboards, higher on G2, G7
and G15 and lower on the other thirteen, mostly fields of the first KAUST competition. With the better of the default and
the tuned configuration, a choice that uses the test set, the placements are identical on twenty-six of the twenty-nine
leaderboards and higher on G2, G7 and G15.

Figure~\ref{fig:finishing} also shows what we would have achieved entering as one team: the method, as well as
its configuration, is chosen on the training data, as the tuned model with the lowest inner cross-validation score, and
only that method is submitted. We would still have won only on 3a-1 $Z_1$, now with NN-GLS, and the placement would drop
on eleven of the other twenty-eight leaderboards, the most on SIC97 and routine SIC2004, consistent with the weak rank
correlations between inner and test scores on the small competitions.

\paragraph{Development-phase additions.} All results come from runs in which every configuration, model variant and
fold design was selected by the rules of the main text (Section~3), without using test scores. During development we
added four methods to cover the families (gradient boosting, NN-GLS, the neural process and the Vecchia GP) and adopted
local context for the foundation models and cross-validation folds shaped like the MODIS cloud gaps; each of these choices is made by those rules, from information available without test values, although the decision to
add them was taken after a first round of test scores had been seen. We also tried a field-level diffusion model on MODIS; five versions were scored on the test set, all far worse than every
reported entry, and we dropped it, which changes no placement. It is not among the reported predictors. Two methods, TabPFN-3.5 and
deepspat, were added after the list of methods had been fixed, the graph networks were run on MODIS, its simulated
version and the KAUST fields only after the other results were known, and the extra information of the original
participants was given to all predictors after runs without it had been scored. The scale of deepspat's learning rates
on SIC97 and SIC2004 was at first set on the SIC97 test set; it is now chosen by inner cross-validation, like the other
settings (Section~\ref{sec:settings}), and the reported entries come from that choice.

Table~\ref{tab:initial} repeats the placements of our best entry with only the methods of the first scoring, and with
those methods and the protocol of the first scoring. Without the later methods, our best placement is lower on eight
leaderboards, because the Vecchia GP is our best entry on six of them (NG1, NG2, G1, G2, G4 and G11), TabPFN-3.5 on
G15 and NN-GLS in routine SIC2004. With the first-scoring protocol as well, it is lower on two more: no 2026 entry scores
below the SPDE approach on MODIS, and the one resolved win, on 3a-1 $Z_1$, disappears. That win therefore rests on a choice made after test scores had
been seen (giving the other variable as a covariate), it does not hold with 25 bootstrap blocks
(Section~\ref{sec:boot}), and its interval is not adjusted for the many comparisons we make. The later additions thus account for
our best placements on these leaderboards; without them, the 2026 entries would fall further behind the original winners than reported in the main text.

\begin{table}[!t]\centering\footnotesize\setlength{\tabcolsep}{2.5pt}
\caption{Sensitivity of the placements to the additions of the development phase (main text, Section~3). Our best entry, its score relative to the original winner and its rank among the published entries: as reported; with only the methods of the first scoring (without gradient boosting, NN-GLS, the neural process, the Vecchia GP, TabPFN-3.5 and deepspat, and without the graph networks on MODIS, its simulated version and KAUST); and with those methods and the protocol of the first scoring (inner folds of $k$-means blocks on MODIS, random context for the foundation models on the large datasets, no extra information on SIC2004 and KAUST2 3a); and, in the last column, with every predictor at its default configuration, also chosen blind to the test set. The coarser DeepKriging enters every column with its configuration tuned on the training data, although it was tuned after the first results were known. `=' marks a tie (main text, Section~3).}
\label{tab:initial}
\resizebox{\textwidth}{!}{\begin{tabular}{lllll}\toprule
Leaderboard & As reported & First-scoring methods & First-scoring methods and protocol & Every predictor at its default \\ \midrule
SIC97 & NN, +1.0\% (2) & NN, +1.0\% (2) & NN, +1.0\% (2) & NN, +0.26\% (2) \\
SIC2004 routine & NN-GLS, +0.10\% (=2) & Trend + DeepKriging, +0.28\% (=3) & Trend + DeepKriging, +0.28\% (=3) & Ordinary kriging, +0.52\% (=4) \\
SIC2004 emergency & Vecchia GP, +1.9\% (2) & TabPFN, +8.8\% (2) & TabPFN, +8.2\% (2) & Vecchia GP, +4.9\% (2) \\
MODIS & GCN-L, $-$8.7\% (1) & TabICL, $-$3.1\% (1) & KED, +1.7\% (2) & TabICL, $-$5.6\% (1) \\
MODIS simulated & NN-GLS, +0.78\% (2) & GP, +0.87\% (2) & GP, +0.87\% (2) & Regression kriging, +1.3\% (2) \\
G1 & Vecchia GP, +0.44\% (6) & Regression kriging, +7.7\% (12) & Regression kriging, +7.7\% (12) & Vecchia GP, +7.6\% (12) \\
G2 & Vecchia GP, +0.02\% (2) & Regression kriging, +0.12\% (9) & Regression kriging, +0.12\% (9) & Vecchia GP, $-$0.006\% (=1) \\
G3 & Regression kriging, +0.25\% (16) & Regression kriging, +0.25\% (16) & Regression kriging, +0.25\% (16) & Regression kriging, +0.26\% (16) \\
G4 & Vecchia GP, +3.7\% (7) & KED, +12.1\% (11) & KED, +12.1\% (11) & Vecchia GP, +13.7\% (11) \\
G5 & KED, +0.02\% (4) & KED, +0.02\% (4) & KED, +0.02\% (4) & Vecchia GP, +0.02\% (5) \\
G6 & KED, +0.16\% (15) & KED, +0.16\% (15) & KED, +0.16\% (15) & KED, +0.15\% (15) \\
G7 & Regression kriging, +0.05\% (6) & Regression kriging, +0.05\% (6) & Regression kriging, +0.05\% (6) & Vecchia GP, $-$0.006\% (=1) \\
G8 & NN-GLS, +0.15\% (15) & GP, +0.15\% (15) & GP, +0.15\% (15) & Ordinary kriging, +0.15\% (15) \\
G9 & GP, +0.02\% (7) & GP, +0.02\% (7) & GP, +0.02\% (7) & Vecchia GP, +0.36\% (16) \\
G10 & GP, +0.004\% (2) & GP, +0.004\% (2) & GP, +0.004\% (2) & GP, +0.02\% (4) \\
G11 & Vecchia GP, +0.03\% (8) & Regression kriging, +0.08\% (12) & Regression kriging, +0.08\% (12) & NN-GLS, +0.13\% (13) \\
G12 & Trend + DeepKriging, +0.11\% (12) & Trend + DeepKriging, +0.11\% (12) & Trend + DeepKriging, +0.11\% (12) & TabPFN-3.5, +0.15\% (13) \\
G13 & GP, +0.10\% (10) & GP, +0.10\% (10) & GP, +0.10\% (10) & GP, +0.15\% (11) \\
G14 & Regression kriging, $-$0.000005\% (=1) & Regression kriging, $-$0.000005\% (=1) & Regression kriging, $-$0.000005\% (=1) & NN-GLS, +0.08\% (16) \\
G15 & TabPFN-3.5, +0.03\% (5) & GP, +0.05\% (10) & GP, +0.05\% (10) & TabPFN-3.5, +0.02\% (4) \\
G16 & GP, $-$0.003\% (=1) & GP, $-$0.003\% (=1) & GP, $-$0.003\% (=1) & Ordinary kriging, +0.05\% (10) \\
NG1 & Vecchia GP, +2.7\% (10) & Ordinary kriging, +6.5\% (13) & Ordinary kriging, +6.5\% (13) & Vecchia GP, +2.7\% (10) \\
NG2 & Vecchia GP, +0.65\% (2) & KED, +12.5\% (11) & KED, +12.5\% (11) & Vecchia GP, +0.66\% (2) \\
1a-1 & Flow matching, +1.0\% (8) & Flow matching, +1.0\% (8) & Flow matching, +1.0\% (8) & TabICL, +1.4\% (8) \\
1a-2 & KED, +1.1\% (7) & KED, +1.1\% (7) & KED, +1.1\% (7) & Regression kriging, +1.2\% (7) \\
3a-1 $Z_1$ & KED, $-$2.0\% (1) & KED, $-$2.0\% (1) & Regression kriging, +0.19\% (3) & KED, $-$2.0\% (1) \\
3a-1 $Z_2$ & Vecchia GP, +0.17\% (3) & GP, +0.30\% (3) & GP, +0.30\% (3) & NN-GLS, +18.2\% (3) \\
3a-2 $Z_1$ & GP, $-$0.0006\% (=1) & GP, $-$0.0006\% (=1) & GP, $-$0.0006\% (=1) & Ordinary kriging, +0.06\% (5) \\
3a-2 $Z_2$ & Trend + DeepKriging, $-$0.03\% (=1) & Trend + DeepKriging, $-$0.03\% (=1) & Trend + DeepKriging, $-$0.03\% (=1) & TabICL, +0.02\% (2) \\
\bottomrule\end{tabular}}\end{table}

\FloatBarrier
\section{Predictive distributions, calibration and computing time}\label{sec:uq}
Each method except inverse distance weighting returns 19 predictive quantiles, at levels 0.05, 0.10, \ldots, 0.95; the
90\% interval runs from the 0.05 to the 0.95 quantile. Each method's predictive quantiles are its own (\emph{native}): from
the kriging variance, a Gaussian process posterior, quantile regression forests, ensemble spread with an estimated noise
variance, generative samples or the foundation models' predictive distributions (Section~\ref{sec:impl}).

\paragraph{MODIS: calibration and computing time.} All our entries run in one environment and were timed on the same
hardware, so their calibration and computing times can be compared directly. \citet{heaton2019} also scored the
predictive distributions of every team with the continuous ranked probability score (CRPS; \citealp{gneiting2007}), the
mean over the test pixels of $\int (F(u)-\mathbf{1}\{y\le u\})^2\,du$ for a predictive distribution function $F$ and
observation $y$, which rewards both calibration and sharpness and is in the units of the data: the lowest CRPS, 0.83\,$^\circ$C, is that of the SPDE approach, whose 95\%
intervals covered 97\% of the test pixels, followed by NNGP (0.85, 95\%) and LatticeKrig (0.87, 96\%). Our CRPS is
approximated from the 19 predictive quantiles, as twice the mean quantile score over the levels 0.05 to 0.95, which
leaves out the tails and so understates the CRPS when observations fall far outside them, as on MODIS, where many observations lie beyond them (see the PIT below); the
lowest is 0.80\,$^\circ$C (TabICL), followed by KED and GCN-L (0.85). The two sets are therefore not directly comparable,
and we read them only as showing that the best 2026 predictive distributions are not clearly better than the best
published ones. Of the published teams, only the three with the lowest CRPS reached the nominal coverage of their 95\% intervals; the
other nine covered 36\% to 93\% of the test pixels. Most of our entries also under-cover their 90\% intervals (below). The teams'
computing times were measured on their own hardware and are not comparable with ours.

\begin{figure}[!t]\centering
\includegraphics[width=0.6\textwidth]{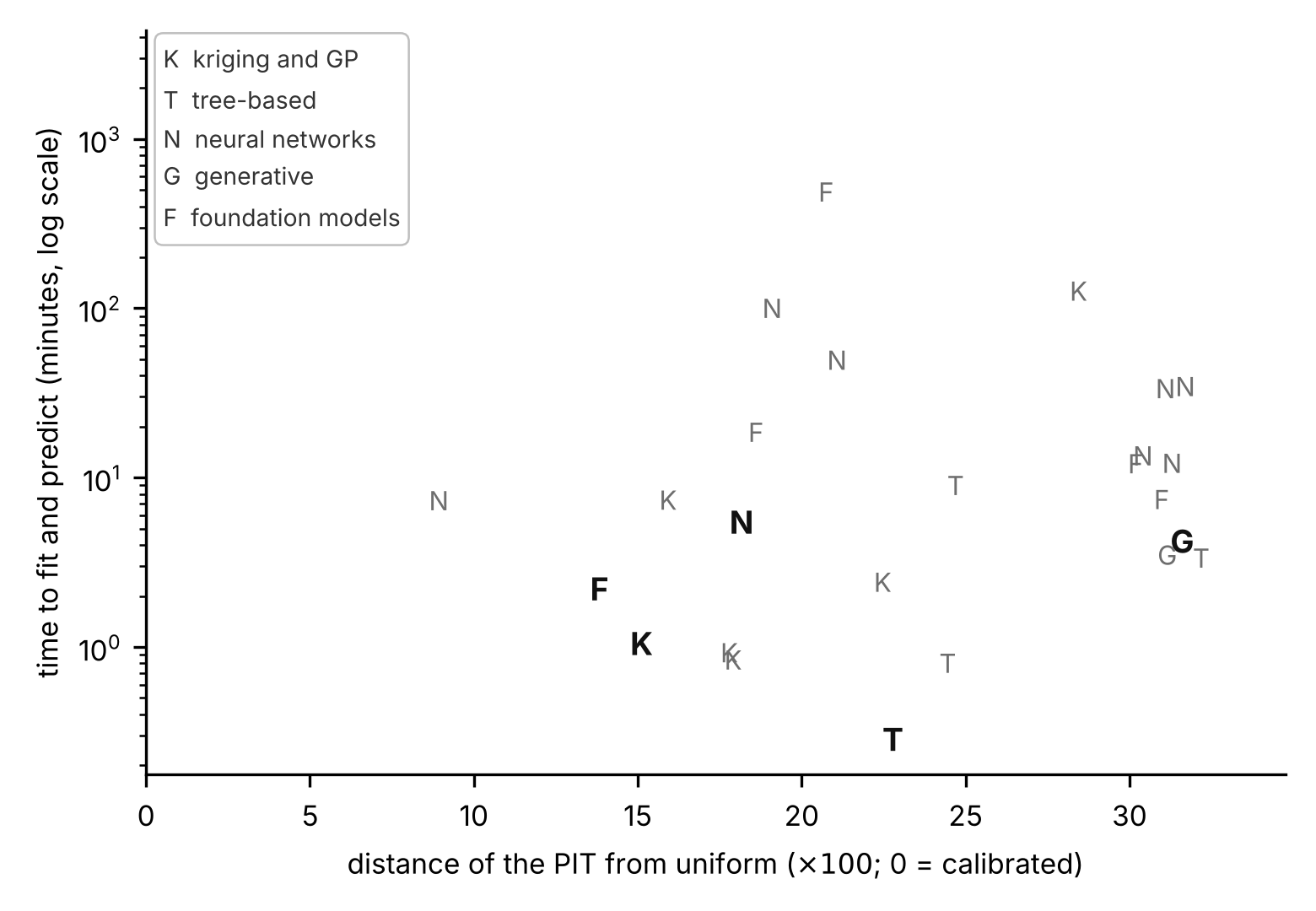}
\caption{Calibration of the predictive distributions against computing time on MODIS, for each tuned 2026 predictor.
The horizontal axis is the area between the distribution function of the PIT over the test pixels and that of the
uniform; the vertical axis is the time to fit the selected configuration and predict, on 16 CPU cores.
Letters are family initials (key in the figure); the vertical axis is in minutes, on a logarithmic scale.
Bold letters mark the best entry of each family, with its test RMSE relative to the SPDE approach, the best published
entry (no winner was named): TabICL (F, $-3.1\%$), KED (K, $-0.7\%$),
NN-GLS (N, $+14.9\%$), RF + kriging (T, $+18.1\%$) and diffusion (G, $+59.3\%$). The graph networks, run on GPUs, are
not shown, among them GCN-L, the entry with the lowest RMSE; inverse distance weighting has no predictive quantiles. The
timing runs are described in Section~\ref{sec:settings} (Computing).\label{fig:uqcost}}
\end{figure}

Figure~\ref{fig:uqcost} places each MODIS predictor by the calibration of its whole predictive distribution and by its
computing time. Calibration is measured with the PIT, the predictive distribution function evaluated at the observed
value, which is uniformly distributed when the predictive distributions are calibrated \citep{gneiting2007calibration};
the calibration score (the distance of the PIT from uniform on the axis of Figure~\ref{fig:uqcost}) is $100\int_0^1|F(u)-u|\,du$, where $F$ is the empirical distribution function of the PIT over the
test pixels, zero only for calibration at every level. From the 19
quantiles, the predictive distribution function is taken as linear between consecutive quantiles and, below the 0.05
and above the 0.95 quantile, as linear with the slope of the outermost interval, clipped to $[0,1]$; observations
beyond the extrapolated tails thus get a PIT of exactly 0 or 1, from 5\% of the test pixels (KED) to 40\% (gradient
boosting), median 21\%, itself a sign of overconfidence. No predictor comes close to calibration: the smallest score,
of the neural process, is 8.9, and the next ones are 13.8 (TabICL) and 15.1 (KED). All underpredict the hot northern gap
(Figure~\ref{fig:modis}), and this bias dominates the score. The three kriging predictors and the Vecchia GP all rank
among the six least miscalibrated of the 28 predictors with predictive quantiles, and their 90\% intervals come close to the nominal coverage, but partly
because they are wide enough to absorb the bias; most other families are both more biased and more overconfident.

The figure also shows that extra computing time does not buy accuracy or calibration: KED and TabICL, among the most
accurate and, after the neural process, the best calibrated, take one to two minutes; the networks take minutes to hours,
and the only better calibrated one, the neural process, is far less accurate; and the slowest predictors, deepspat and TabPFN-3.5 (two and eight hours), are neither more accurate nor
better calibrated. These times are for a single fit with the selected configuration. Tuning repeats such a fit many
times, for every candidate configuration and every inner cross-validation fold, so it accounts for most of the computing
cost of every tuned entry.

\begin{figure}[!ht]\centering
\includegraphics[width=\textwidth,height=0.6\textheight,keepaspectratio]{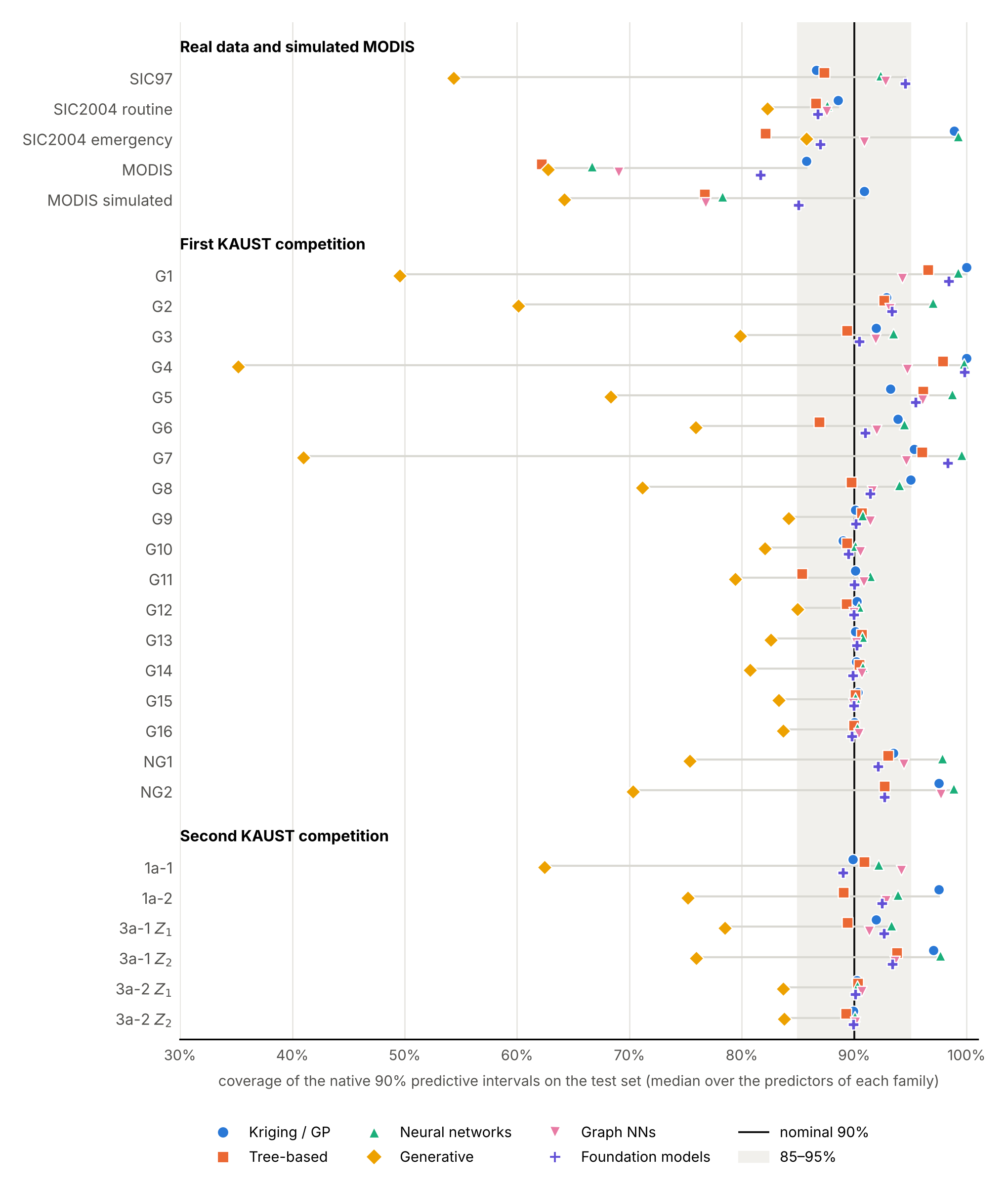}
\caption{Coverage of the native 90\% predictive intervals on the competition test sets: for each leaderboard, the median
over the predictors of each family, one marker per family as in Figure~\ref{fig:leaderboards_all}; the line marks the
nominal 90\% and the band 85--95\%. Baselines (inverse distance
weighting has no intervals) are omitted. The coverage of every predictor is in \texttt{results\_best/<key>/uq\_metrics.csv}
in the repository.\label{fig:coverage}}
\end{figure}

\paragraph{All leaderboards.} Figure~\ref{fig:coverage}
gives the coverage of the native 90\% intervals on each leaderboard, by family. On MODIS, where the test pixels form large gaps, every family under-covers, kriging and Gaussian processes least; on its
simulated version, every family except kriging and Gaussian processes. On the other twenty-seven leaderboards, where the test
locations are interspersed with the training locations, most families are close to the nominal level (the median over
these leaderboards of each family's coverage is 90--93\%). The generative models under-cover (median 80\%, as low as 15\%), and so
does deepspat on SIC97 and routine SIC2004 (57\% and 63\%), which the family medians of the figure do not show. On three of the smooth fields
without a nugget (G1, G4 and G7) the intervals of many predictors over-cover, several reaching full coverage, and some
families also over-cover on the SIC2004 emergency, NG2, 1a-2 and 3a-1 $Z_2$.

\FloatBarrier
\section{Implementation details}\label{sec:impl}
\subsection{The methods}\label{sec:families}
The \emph{coordinate networks} are NN, Trend + NN and NN-cov, fully connected networks on the coordinates. Labels: ``-cov'' marks a predictor given the optional covariate as an input; ``Trend +'' a network or kriging fitted to
the residuals of a least-squares trend, the drift of KED (linear in the coordinates, plus the covariate where there is
one; on SIC97, linear in elevation only); ``RF + kriging'' the random forest with kriged residuals; NN, DeepKriging
and DeepKriging-P are described below, and the predictor labels are those of the main text (Table~2). The published MODIS
entries MRA, LAGP and FRK are the multi-resolution approximation, local approximate Gaussian processes and fixed rank
kriging \citep{heaton2019}. The context of the foundation models and their local version are defined in
Section~\ref{sec:settings} (Foundation models).

\emph{Kriging and Gaussian processes.} Kriging assumes a stationary and isotropic Gaussian random field, whose
variogram is fitted by weighted least squares \citep{cressie1993}. Ordinary kriging assumes a constant mean; regression kriging fits a global trend by least squares and kriges its residuals;
kriging with external drift (KED, universal kriging) instead estimates the trend by generalized least
squares, accounting for the spatial correlation of the errors. The Gaussian process (GP) has an anisotropic Mat\'ern
covariance fitted by maximum likelihood. The Vecchia Gaussian process (Vecchia GP) fits the same kind of covariance on all
training points through the Vecchia approximation of the likelihood. The deep compositional spatial model (deepspat), a
nonstationary Gaussian process, warps the domain with a composition of learned functions.

\emph{Tree-based methods.} The random forest \citep{breiman2001} splits on the coordinates and any covariate; RFsp
\citep{hengl2018} uses the distances to the observation locations as features instead of the coordinates.
Gradient-boosted trees (Gradient boosting; LightGBM, \citealp{friedman2001gbm,ke2017lightgbm}) receive, in addition, rotated
copies of the coordinates, so that splits are not limited to the east--west and north--south directions.

\emph{Neural networks.} The neural network (NN) is a fully connected network with the coordinates as inputs (NN-cov adds the covariate). DeepKriging \citep{chen2024deepkriging} feeds a
network with the values at each location of a multi-resolution basis of compactly supported Wendland functions, either
with the authors' basis and architecture (DeepKriging-P) or with a coarser basis. NN-GLS \citep{zhan2025nngls} is a hybrid: a network for the mean, trained with a
generalized-least-squares loss under a nearest-neighbor Gaussian process covariance, whose residuals are then
kriged, so that it is close to regression kriging with a network trend; its results partly reflect its covariance model,
so they say little about networks without one. The neural process \citep{garnelo2018np,kim2019anp,nguyen2022tnp} predicts
each location from its nearest observations and is meta-trained on the training data.

\emph{Generative models.} Denoising diffusion \citep{ho2020,han2022card} and flow matching \citep{lipman2023} learn to
draw samples from a distribution by transforming Gaussian noise. Diffusion trains a network to remove noise added to
the data in many small steps, and generates by reversing those steps; flow matching trains a network to move noise
continuously toward the data. Here both model the distribution of the value at a single location given that
location's inputs (the same basis functions as DeepKriging, and any covariate), separately for each location, and the
prediction is the mean of the draws.

\emph{Graph neural networks.} Graph convolutional networks \citep{kipf2017} act on a graph linking each location to its
nearest neighbors. In GCN the input at each node is its coordinates (and covariates) only; in GCN-L each node also
receives the observed value as an extra input, set to zero, with a flag, at the test locations and, during training, at a
random half of the training locations, so that the network learns to predict a hidden value from the observed values of
its neighbors.

\emph{Foundation models.} A foundation model is a large network trained once, in advance, on a vast collection of
data and then applied to new problems without being refitted. The tabular foundation models TabPFN v2 (TabPFN;
\citealp{hollmann2025}), TabPFN-3.5 \citep{priorlabs2026tabpfn35} and TabICL v2 (TabICL; \citealp{qu2025tabicl,qu2026tabiclv2})
were pretrained on large collections of tabular datasets; given the training data of a new problem and the
test locations, they return the predictions in a single pass, without estimating any parameters on that problem. The
training points a prediction conditions on, its conditioning set, are called the \emph{context}. The GeoCLIP \citep{vivanco2023geoclip} and SatCLIP \citep{klemmer2023satclip} location encoders are followed by a network.

\subsection{Settings of each method}\label{sec:settings}
All methods are implemented in Python, except deepspat, the deep compositional spatial model, which runs in R through the
\texttt{deepspat} package; the code and per-fold predictions of all reported predictors are in the repository
described in the Data availability statement of the main text, except the predictions on KAUST2 1b and 3b, whose summaries the repository holds.

\paragraph{Kriging.} By default, a spherical model (the family is tuned) is fitted by weighted least squares, with
weights $N_j/h_j^2$ (as in \texttt{gstat}; \citealp{pebesma2004}), to an empirical variogram with 15 lags up to one third of
the diagonal of the bounding box of the data. Ordinary kriging assumes a constant mean, regression kriging kriges the
residuals of an ordinary least-squares trend, and kriging with external drift estimates the trend jointly; where a
leaderboard has no covariate, the trend of the last two is linear in the coordinates, so that KED there is universal
kriging. On the large datasets all kriging is local, with the 64 nearest neighbors by default. When there are
more than 1{,}000 training points (MODIS, its simulated version and the KAUST fields), the empirical variogram is computed differently: with log-spaced
lags, using every pair of points within 3.2 distance units from the full data and the longer lags from a
3{,}000-point subsample, up to a fixed cutoff of 60 units (kilometers on MODIS; on the KAUST fields, the unit square is scaled by
100), with the pair counts in the weights capped at 5{,}000. The number of lags and the cutoff fraction in the search space are not used in this case, so on these datasets the
tuning of kriging is effectively over the variogram family and the neighborhood size. The Gaussian process
uses an anisotropic Mat\'ern covariance with nugget (smoothness $\nu=3/2$ by default; tuned), hyperparameters estimated by
maximum marginal likelihood on a subsample (1{,}000 points by default; tuned) and prediction from the nearest
neighbors on the large datasets (128 by default; tuned).

\paragraph{Vecchia Gaussian process.} After a linear trend in the coordinates (and covariates, where there are any) is removed by least squares (by default;
whether to remove it is tuned), the residuals are modeled by an anisotropic Mat\'ern covariance with nugget, with smoothness
$\nu\in\{0.5,1.5,2.5\}$ fixed per configuration (1.5 by default). Its sill, two range parameters and nugget are estimated
by maximizing the Vecchia approximation of the likelihood \citep{vecchia1988,katzfuss2021vecchia} on all training
points, as in GpGp \citep{guinness2018}: the points are ordered coarse to fine (one point per cell of
successively finer grids, an approximation of the max-min ordering), each point is conditioned on its $m$ nearest
previous points ($m=30$ by default), and the likelihood is optimized by L-BFGS with automatic differentiation. Prediction
is simple kriging of the residuals from the $m_{\text{pred}}$ nearest training points under the fitted covariance (60 by
default), added to the trend; the kriging variance is the native uncertainty.

\paragraph{deepspat.} We use the
function \texttt{deepspat\_nn\_GP} of the R package \texttt{deepspat} 0.3.5 of \citet{zammitmangion2022}. A linear trend in the
coordinates (and covariates, where there are any) is first removed by least squares, and the coordinates are rescaled to
$[-0.5,0.5]^2$. The domain is warped by the composition of two axial warping units (\texttt{AWU}, $r=50$, one per
coordinate), a radial basis function block (\texttt{RBF\_block(1)}) and a M\"obius transformation (\texttt{LFT}); the
residual field on the warped domain is a nearest-neighbor Gaussian process with exponential covariance (family
\texttt{exp\_nonstat}) and $m=25$ neighbors under a max-min ordering, fitted by restricted maximum likelihood with 150
optimization steps. Prediction uses the 25 nearest training points, and the native uncertainty is the predictive variance
including measurement error. It is tuned like the other methods: the default and 19 random configurations over the
warping architecture (with or without the M\"obius transformation, or with a second radial basis block), the axial basis
size 25, 50 or 100, $m=15$ or 25 neighbors and 150 or 300 optimization steps, each scored by five-fold cross-validation on
the training data. The scale of its learning rates is also chosen by inner cross-validation: the package defaults or the
defaults scaled by 0.1 on SIC97 and SIC2004, and the defaults scaled by 0.1 or by 0.03 on the large datasets; a scale at
which training diverges or does not complete within the 20-hour cap is not a candidate. In its tuning runs on the eighteen
leaderboards where it fitted (twenty-three runs, as five leaderboards were tuned with and without the covariate), some
configurations were stopped by the 20-hour cap (72 of 460) or failed numerically (18, all on SIC97), and the selected
configuration has a lower inner cross-validation score than the default in twenty of the twenty-three runs. A cross-validation fold that failed in training was refitted at a
learning rate ten times smaller where the selected scale allowed it (on SIC97 and SIC2004 only), and reported as failed if
it failed again. deepspat failed to fit on eleven leaderboards, the SIC2004 emergency, 1a-1, 1a-2, 3a-1 $Z_2$, NG1, NG2,
G1, G2, G4, G5 and G7: restricted maximum likelihood drives the nugget toward zero and the single-precision
nearest-neighbor covariance becomes singular at every learning rate tried. On KAUST1 it thus fitted on eleven of the
sixteen 1b fields but failed on the other five and on both 2a fields. It also failed on all eight KAUST2 1b and 3b
leaderboards. We report these as failures and did not modify the package.

\paragraph{Tree-based methods.} Random forests with 500 trees (200 on the large datasets) on coordinates and covariates; RFsp uses the distances to the
observation locations (to 100 $k$-means reference points on the large datasets) and the covariates as features, without the
coordinates \citep{hengl2018}; predictive quantiles from quantile regression forests \citep{meinshausen2006}. The random forest
with kriged residuals uses a Gaussian predictive distribution from the kriging variance of its out-of-bag residuals. Gradient boosting uses LightGBM \citep{ke2017lightgbm}
with 500 trees, learning rate 0.05, 31 leaves and row subsampling 0.8, on the coordinates, their projections on further directions (at 45, 90 and 135 degrees by default, so that the
90-degree projection repeats the second coordinate) and the covariates; its predictive quantiles come from one quantile-loss model per level. The minimum leaf size is capped at
one tenth of the training set, which binds on SIC97.

\paragraph{Neural networks.} Each coordinate network is the average of five fully connected neural networks (three hidden layers of 64, 64
and 32 units, ReLU, dropout 0.1, weight decay $10^{-3}$, AdamW with learning rate $3\times10^{-3}$, early stopping with
patience 40 checks every 5 epochs); on the large datasets they are trained in mini-batches of 512 with learning rate $10^{-3}$
(tuned over $3\times10^{-4}$, $10^{-3}$ and $3\times10^{-3}$, in place of the full-batch learning rate), at most 300 epochs and patience 15. DeepKriging-P follows the published specification \citep[Supplement S2 of][]{chen2024deepkriging}: a
multi-resolution Wendland basis \citep{wendland1995} with $(9\cdot2^{h-1}+1)^2$ knots at level $h$ and
$H=1+\lceil\log_2(\sqrt{N}/10)\rceil$ levels, clipped to between 3 and 5 (the bounds are tuned), bandwidth 2.5 times the knot spacing, a three-layer network of width 100 with
batch normalization and dropout 0.5, trained with Adam for 200 epochs in batches of 64, five members. Native uncertainty
combines ensemble spread with a validation-based noise variance. Our DeepKriging uses a coarser basis, three levels with
5, 10 and 15 knots per side (8, 16 and 32 on the large datasets) and the same bandwidth rule, fed to the coordinate
network described above, but with dropout 0.2 and weight decay $10^{-2}$ by default.
NN-GLS follows \citet{zhan2025nngls} with three departures: where it has no covariate (MODIS and its simulated version, 1a, the first KAUST
competition, and the leaderboards on which inner cross-validation dropped the covariate) the network takes the coordinates, so that space enters the mean as well as the error, as in their coordinate-input comparator;
the covariance parameters come from a weighted-least-squares variogram fit rather than from the likelihood; and the
Vecchia ordering is random. The mean is a fully connected neural network (two hidden layers of 32 and 16 units, weight
decay $10^{-2}$) of the covariates, or of the coordinates when a dataset has none; it is trained with a
generalized-least-squares loss in which the residuals are decorrelated by a nearest-neighbor (Vecchia) approximation with 15
neighbors of an exponential covariance, estimated first from the residuals of a linear trend and then from the network
residuals, over two rounds; predictions add kriging of the network residuals, whose variance is the native uncertainty.
The neural process is a local transformer neural process in the spirit of \citet{kim2019anp} and \citet{nguyen2022tnp},
without a latent path: for each target, its 32 nearest observations (offset, distance, value and covariates) and a query
token are encoded jointly by a two-layer transformer (width 64, four heads), which outputs a Gaussian mean and variance
from the query token; it is meta-trained for up to 4{,}000 steps with early stopping by predicting each training
point from its nearest other training points with the Gaussian negative log-likelihood, three members.

\paragraph{Generative models.} A conditional denoising diffusion model in the CARD style \citep{ho2020,han2022card} with
$T=100$ steps and 50 samples per location, sampling through the predicted noise-free value clipped to $\pm5$ training standard
deviations; and conditional flow matching \citep{lipman2023} integrated with 20 Euler steps, 50 samples per location.

\paragraph{Graph neural networks.} Graph convolutional networks \citep{kipf2017} on an 8-nearest-neighbor graph, four layers of five propagation hops each, three-network ensembles.
On MODIS, its simulated version and the KAUST fields, where a fit takes hours on a CPU, they were run on GPUs.

\paragraph{Foundation models.} TabPFN v2 \citep{hollmann2025} and TabICL v2 \citep{qu2025tabicl,qu2026tabiclv2} with their released
checkpoints, on coordinates and covariates. When the training set is larger than the context (10{,}000 rows for TabPFN
and 20{,}000 for TabICL by default, that is, on MODIS, its simulated version and the KAUST fields), the context is local: the test points are
split by $k$-means into spatially compact groups of about 4{,}000 points, and each group is predicted from the training
points closest to any of its members, as many as the context size; a separate in-context fit is made for each group. On these datasets each foundation model is tuned twice,
with local and with random context, and inner cross-validation chooses between the two. With its default settings on MODIS,
TabICL scores an RMSE of 1.444\,$^\circ$C with the local context and 1.959\,$^\circ$C with a random 20{,}000-row context. The number of ensemble members and the
context size are tuned (Section~\ref{sec:tune}); on MODIS TabICL selected a 5{,}000-point context.
TabPFN-3.5 \citep{priorlabs2026tabpfn35} uses the released
weights (Hugging Face repository \texttt{Prior-Labs/tabpfn\_3\_5}, TABPFN-3.5 License v1.0, which does not permit commercial use) through the \texttt{tabpfn} package, run locally on one GPU. It is used as TabPFN v2, with local context on the
large datasets (tuned also with random context, as above, except on the four 3a leaderboards) and 8 ensemble members by default, but with a default context of 50{,}000 rows; on the 3a fields, with 45{,}000
training points, the default context is the whole training set. These defaults and the search space were fixed before
any test score of TabPFN-3.5 was seen. It is tuned over the number of ensemble members
$\in\{4,8,16,32\}$ and the context size $\in\{10{,}000,25{,}000,50{,}000\}$ (12 distinct configurations, the default
among them; 4 on the small datasets, where the context exceeds the training set). GeoCLIP
\citep{vivanco2023geoclip} and SatCLIP \citep{klemmer2023satclip} embeddings of longitude and latitude are reduced to 32
principal components and fed to the average of several fully connected neural networks.

\paragraph{Computing.} Most default runs on the ten leaderboards of SIC97, SIC2004, MODIS and the second KAUST
competition used one to four CPU threads in a CPU-only container (4 virtual CPUs); those of the other leaderboards, the
tuning and the DeepKriging ensembles on the large datasets ran on CPU nodes of a university computing cluster with 16
cores per job. Gradient boosting, NN-GLS, the neural process and the Vecchia Gaussian process, and the local-context runs
of TabPFN and TabICL ran, including their default configurations, on nodes of the same cluster with 4 or 16 cores per
job. deepspat ran in R (through the TensorFlow back end of \texttt{deepspat}) on nodes of the same cluster: most of its
tuning on CPU nodes with 16 cores per job, and part of its runs on the large datasets on one GPU with 4 cores (NVIDIA
A10, A40 or RTX 6000). TabPFN-3.5 ran on one GPU; on
MODIS its selected configuration takes 709\,s there, against 8.1 hours on the 16 CPU cores of Figure~\ref{fig:uqcost}.
The graph networks ran on one GPU each (NVIDIA A10 or A40) on MODIS, its simulated version and the KAUST fields. In all, the study used about
300{,}000 core-hours and 1{,}250 GPU-hours on the cluster, of which about 280{,}000 and 1{,}230 were for the twenty-nine
leaderboards, reconstructed from the job logs.
The times in Figure~\ref{fig:uqcost} come instead from separate timing runs on MODIS: each predictor's selected
configuration was refitted once on all training pixels on the same node type (16 CPU cores, no GPU), and, for the
26 predictors refitted (inverse distance weighting and the 25 tuned predictors shown in Figure~\ref{fig:uqcost}, that is,
all but the graph networks, which ran on GPUs, and the linear trend), the refits give the stored test RMSE identically for 23; the other three are not fully seeded and differ from it by
0.046 (GeoCLIP), 0.026 (NN-GLS) and 0.001\,$^\circ$C (TabPFN-3.5), at most 3\% of the RMSE. (The 28 predictors
with predictive quantiles in Section~\ref{sec:uq} are these 25, the two graph networks and the linear trend.) Times are comparable only between entries run on the same hardware.

\providecommand{\bibcommenthead}{}\providecommand{\urlprefix}{}

\FloatBarrier

\section{The precipitation pilot competition}\label{sec:cerra}
The pilot of the main text (Section~5, Figure~4) was run with a procedure frozen in a design document, together with
four checks of its conclusions, before any field of 2020 or later was opened: every model was fitted on archives ending
in 2019, its choices were made on validation days of 2020, and its predictions were scored on the wet days of
2021--2025, whose fields had been downloaded but not opened (\texttt{explorations/cerra/confirmation/}, design in
\texttt{DESIGN.md}, record in \texttt{FACTS.md}). The design committed to reporting the results whatever they showed.

\paragraph{Data and tasks.} Daily total precipitation (mm, 24 hours ending at 06 UTC) from the CERRA-Land reanalysis
\citep{cerraland}, on its 5.5\,km grid, in a 74 by 117 crop over the Alps (about 44 to 47.6$^\circ$N, 6 to
14.4$^\circ$E), land cells only (land-sea mask at least 0.5; 7{,}670 cells), with the CERRA orography as elevation. The
test days are all 983 days of 2021--2025 with land-mean precipitation above 2\,mm; the validation days, used for every
choice, are the first two days of each month of 2020 on which land-mean precipitation exceeded 2\,mm (24 days), few
enough that the choices made on them (shrinkage factors, the reconstruction from empirical orthogonal functions,
references and diffusion checkpoints) are themselves uncertain; the data of 2026 and later were never read. On each day,
a random network of land cells is observed, 2\% (153 cells, a mean spacing of about 40\,km, like a rain-gauge network) or
0.5\% (38 cells), drawn with a fixed seed per day and task, and every other land cell is predicted. The entrants may train
on an archive of past daily fields: 3 years (2017--2019), 10 years (2010--2019) or 30 years (1990--2019).

\paragraph{Methods.} (i) Regression kriging on the day alone: a linear trend on elevation and a Mat\'ern Gaussian
process on the residuals (range, smoothness between 0.2 and 2.5, and nugget estimated by maximum likelihood on the
observed cells), with prediction from the 150 nearest
observed cells, that is, all of them with 0.5\% observed and all but three with 2\%. (ii) Simple kriging with the archive covariance: the archive mean and empirical covariance at the land
cells, over all days of the archive (dry ones included), shrunk toward the diagonal by a factor chosen on the validation
days. (iii) Reconstruction from the leading
empirical orthogonal functions (EOFs) of the archive, with the number of functions and a ridge penalty chosen on the validation
days. (iv) Simple kriging of $\log(1+y)$ with the archive covariance of the transformed fields. (v) Kriging after a
censored-Gaussian transformation: at each cell the archive gives the probability of a dry day (below 0.1\,mm) and a
gamma distribution truncated at 0.1\,mm for the wet amounts, which map the rain to a latent Gaussian score; the
correlation of the latent scores comes from the archive, shrunk toward the identity by a factor chosen on the
validation days, and the observed dry cells enter as censored values, sampled by Gibbs sampling (univariate truncated
normals, 100 burn-in sweeps, then 32 draws every 5 sweeps). (vi) TabICL
v2 \citep{qu2026tabiclv2} with the coordinates, elevation, archive mean and standard deviation and the loadings on the ten
leading empirical orthogonal functions as features, and the day's observed cells as context (package defaults, \texttt{random\_state} 0); and (vii) the same with
the prediction of (ii) added as a feature. (viii) A conditional denoising diffusion model \citep{ho2020}, a U-Net that
generates the whole field in $\log(1+y)$, conditioned on the observed cells, the observation mask, elevation and the
archive simple-kriging prediction of (ii), but with the shrinkage factor fixed at 0.1 in the design rather than the one
chosen for (ii) on the validation days (0.01 to 0.03); the U-Net has three levels, 64 base channels and a sinusoidal embedding of the 1{,}000 noise
steps, of which the sampler uses 250, with the predicted clean field clipped at the
99.99\% quantile of the training archive's land values on the transformed scale. It is trained on the archive with random sparse networks of both densities (AdamW, learning rate
$2\times10^{-4}$, weight decay $10^{-4}$, batches of 16 days, an exponential moving average of the weights, 300 passes
over the archive); the checkpoint with the lowest validation RMSE of the mean of 8 samples is kept, as fixed in the
design, and 32 samples are drawn per test day; its point prediction is their mean. For the
scores of predictive distributions each method except (iii) gives 32 values per cell (the quantiles at levels $(k - 0.5)/32$, $k=1,\ldots,32$, for
the Gaussian and TabICL predictive distributions, the samples for diffusion and for (v)), clipped at zero; the
probability of exceeding a threshold is the fraction of the 32 values above it.

\paragraph{Scores and references.} RMSE, the CRPS and the Brier score for exceeding 10 and 25\,mm, over all predicted
cells of the 983 test days. The CRPS is the fair ensemble estimator for the methods whose 32 values are samples,
diffusion and (v), and the average quantile score over the 32 levels,
$\frac{2}{32}\sum_k (\mathbf{1}\{y<q_k\}-\tau_k)(q_k-y)$ with $\tau_k=(k-0.5)/32$, for those whose 32 values are
quantiles; the fair estimator would understate the CRPS of the latter by about 3\%. The primary intervals, fixed in the design, come
from a paired bootstrap of whole test days: each of the 10{,}000 resamples draws 983 test days with replacement, and a
method and the reference are scored on the same resample, so the interval is that of their ratio and allows for the
dependence of the errors within a day. A method beats the reference if the upper 95\% limit is below one, and the reference
is better if the lower limit is above one. The secondary intervals, also fixed in the design, come from a block
bootstrap: the predicted cells of each day are grouped into 10 blocks by $k$-means on their coordinates (9{,}830
blocks per task), resampled with replacement, and a method's score on the resample is divided by the reference's score
on the test set. They ignore the pairing and are wider. The references are chosen on the validation days: for the RMSE,
the best of (i), (ii) and (v), which was (ii) for every archive and task; for the CRPS and Brier scores, the better of
(iv) and (v), which was (v) throughout. The references are kriging methods, so the reconstruction from empirical
orthogonal functions, (iii), is not among them; for the probabilistic scores only the two kriging methods built for
skewed, zero-inflated rain, on the log scale and after the censored-Gaussian transformation, are candidates, because the
Gaussian predictive distributions of (i) and (ii) put mass below zero.

\paragraph{Checks fixed in advance.} The design stated four expected conclusions and how each would be judged,
under the primary intervals: (C1) diffusion beats the CRPS reference with 10 and 30 years at both densities; (C2) on the
RMSE, diffusion is at most narrowly ahead of the reference, reported as observed; (C3) the zero-shot diffusion models
have a higher RMSE than kriging on the day alone, (i), with their CRPS reported as observed; (C4) TabICL on the day alone
lowers the CRPS relative to (i) with 2\% observed. C1, C3 and C4 are confirmed; C2 is reported under Results.

\paragraph{Deviations and incidents.} Three deviations were recorded before any test score was computed. The
code loads whole daily fields and writes the withheld values during prediction, so the rule that withheld values are read
only for scoring was met procedurally: those files were made read-only and opened by nothing until every prediction had
been written (D1). The design misdescribed the seed of the observation networks, which follows the rule of the
code unchanged (D2). Instead of committing the 70\,GB of predictions before scoring, their SHA-256 checksums were committed,
and the scoring script checked every file against them (D3). In addition, a full disk stopped some jobs before any test
score was computed: the two diffusion models with 30 years of history (with and without the input (ii)) were retrained
from scratch with unchanged code and seed, where the design allowed a restart, with the next seed, only after a
divergence; the Carpathian zero-shot model keeps the best of its 12 checkpoints, saved before the failure; and three runs
of (v) were rerun unchanged. No method failed or was dropped.

\paragraph{Zero-shot.} For prediction without the history of the region, diffusion models with the same architecture,
but without the input (ii), were trained on 30 years (1990--2019) of two 74 by 117 windows of the Carpathians and the
Pyrenees, chosen from the land mask and orography alone to resemble the Alps, and applied to the Alpine test days; the
checkpoint of each was chosen on validation days of that region (2020), so no Alpine data entered their training,
although the sampler setting, fixed in the design, had been chosen earlier on validation days of all three regions, the
Alps included.
Kriging on the day alone was also run with borrowed covariance parameters, the medians of maximum-likelihood fits to 48
days of 2016--2019 of each region, and TabICL was run on the day's observed cells alone (coordinates and elevation),
which needs no archive and is zero-shot by construction. The reference is regression kriging on the day alone, (i), as a
model without Alpine history has no archive.

\paragraph{Results.} Tables~\ref{tab:cerra} and \ref{tab:cerrabrier} give the ratio of every method to the reference,
and Table~\ref{tab:cerrazero} those of the zero-shot models. With 10 or 30 years of history, diffusion beats the reference
on the CRPS and on both Brier scores at both densities (C1); with 3 years it beats it with 2\% observed, and with 0.5\%
observed it is behind on the CRPS and not separated on the Brier scores. On the RMSE it is behind archive kriging with 3
years at both densities and with 10 years and 0.5\% observed, and ahead with 10 years and 2\% observed and with 30 years
at both densities, by 2 to 4\% where it is ahead (C2). TabICL is behind archive kriging on the RMSE in every case and
behind the reference on every score with 0.5\% observed; with 2\% observed it beats the reference on the CRPS only with
3 years, and on the Brier scores in most cases, in all of them with the archive kriging prediction as a feature.
Without Alpine history, diffusion is worse than kriging on the day alone in RMSE (5 to 13\%; C3), while its CRPS is 2
to 7\% lower in three of the four cases (all but the Carpathian model with 0.5\% observed); on the Brier scores the
Pyrenean model is better for 10\,mm and worse for 25\,mm, and the Carpathian model worse for both. Borrowing the
covariance of another region changes kriging on the day alone by at most 3\%. TabICL on the day alone lowers the CRPS by
8\% with 2\% observed (C4) but not with 0.5\%, and it is worse on the RMSE at both densities. The same diffusion model
trained on the Alps' own 30 years, also without the input (ii), is 17 to 34\% better than kriging on the day alone on
every score, so the loss comes from the change of region, not from the model. Under the secondary intervals
(\texttt{paper/ADDITIONAL\_RESULTS.md}) 45 of the differences resolved under the primary ones are no longer resolved,
among them the RMSE of diffusion with 3 years and 2\% observed and with 10 and 30 years and 0.5\% observed, and the CRPS
of diffusion with 3 years and 0.5\% observed; no difference changes direction.

\begin{table}[!t]\centering\footnotesize\setlength{\tabcolsep}{3pt}
\caption{Precipitation pilot competition, all 983 test days of 2021--2025: RMSE and CRPS of every method relative to the reference chosen on the 2020 validation days (RMSE: archive kriging (ii); CRPS: censored-Gaussian kriging (v)), by the length of the archive (past fields), with the 95\% intervals of the design (test days resampled whole, method and reference on the same resample). Bold: the interval excludes one (decided at full precision, so a limit printed as 1.000 can be bold); a limit that rounds to 1.00 is given to three decimals.}\label{tab:cerra}
\resizebox{\textwidth}{!}{\begin{tabular}{lllllll}\toprule
 & \multicolumn{3}{c}{2\% observed} & \multicolumn{3}{c}{0.5\% observed} \\ \cmidrule(lr){2-4}\cmidrule(lr){5-7}
Method & 3 years & 10 years & 30 years & 3 years & 10 years & 30 years \\ \midrule
\multicolumn{7}{l}{\emph{RMSE}} \\
(i) Kriging, day alone & \textbf{1.15 [1.14, 1.17]} & \textbf{1.24 [1.22, 1.25]} & \textbf{1.28 [1.27, 1.30]} & \textbf{1.19 [1.17, 1.21]} & \textbf{1.22 [1.20, 1.24]} & \textbf{1.24 [1.22, 1.26]} \\
(ii) Archive kriging & reference & reference & reference & reference & reference & reference \\
(iii) EOF reconstruction & \textbf{1.01 [1.01, 1.01]} & \textbf{1.01 [1.01, 1.01]} & \textbf{1.01 [1.01, 1.01]} & \textbf{1.01 [1.01, 1.02]} & \textbf{1.01 [1.005, 1.01]} & \textbf{1.01 [1.004, 1.01]} \\
(v) Censored-Gaussian kriging & \textbf{1.02 [1.02, 1.03]} & \textbf{1.03 [1.03, 1.04]} & \textbf{1.02 [1.02, 1.03]} & \textbf{1.01 [1.002, 1.02]} & \textbf{1.02 [1.01, 1.03]} & \textbf{1.01 [1.01, 1.02]} \\
(vi) TabICL & \textbf{1.02 [1.02, 1.03]} & \textbf{1.08 [1.07, 1.09]} & \textbf{1.11 [1.10, 1.12]} & \textbf{1.12 [1.11, 1.14]} & \textbf{1.14 [1.13, 1.16]} & \textbf{1.15 [1.13, 1.17]} \\
(vii) TabICL + kriging & \textbf{1.01 [1.005, 1.02]} & \textbf{1.05 [1.04, 1.05]} & \textbf{1.06 [1.06, 1.07]} & \textbf{1.13 [1.12, 1.15]} & \textbf{1.15 [1.14, 1.17]} & \textbf{1.16 [1.15, 1.18]} \\
(viii) Diffusion & \textbf{1.02 [1.01, 1.02]} & \textbf{0.98 [0.97, 0.98]} & \textbf{0.96 [0.95, 0.96]} & \textbf{1.03 [1.02, 1.04]} & \textbf{1.02 [1.01, 1.02]} & \textbf{0.98 [0.98, 0.99]} \\
\midrule
\multicolumn{7}{l}{\emph{CRPS}} \\
(i) Kriging, day alone & \textbf{1.24 [1.23, 1.25]} & \textbf{1.32 [1.31, 1.33]} & \textbf{1.38 [1.37, 1.39]} & \textbf{1.31 [1.29, 1.33]} & \textbf{1.33 [1.32, 1.35]} & \textbf{1.36 [1.35, 1.38]} \\
(ii) Archive kriging & \textbf{1.11 [1.10, 1.12]} & \textbf{1.08 [1.07, 1.08]} & \textbf{1.07 [1.06, 1.08]} & \textbf{1.11 [1.10, 1.12]} & \textbf{1.10 [1.09, 1.10]} & \textbf{1.10 [1.09, 1.11]} \\
(iv) Log-scale archive kriging & \textbf{1.10 [1.09, 1.11]} & \textbf{1.08 [1.07, 1.08]} & \textbf{1.07 [1.06, 1.07]} & \textbf{1.07 [1.06, 1.08]} & \textbf{1.06 [1.05, 1.06]} & \textbf{1.06 [1.05, 1.06]} \\
(v) Censored-Gaussian kriging & reference & reference & reference & reference & reference & reference \\
(vi) TabICL & \textbf{0.98 [0.98, 0.99]} & \textbf{1.03 [1.02, 1.03]} & \textbf{1.06 [1.06, 1.07]} & \textbf{1.14 [1.13, 1.16]} & \textbf{1.15 [1.14, 1.17]} & \textbf{1.17 [1.16, 1.19]} \\
(vii) TabICL + kriging & \textbf{0.98 [0.97, 0.98]} & 1.00 [0.996, 1.01] & \textbf{1.02 [1.01, 1.03]} & \textbf{1.16 [1.15, 1.17]} & \textbf{1.17 [1.16, 1.19]} & \textbf{1.20 [1.18, 1.21]} \\
(viii) Diffusion & \textbf{0.97 [0.96, 0.97]} & \textbf{0.90 [0.90, 0.91]} & \textbf{0.88 [0.87, 0.89]} & \textbf{1.01 [1.002, 1.02]} & \textbf{0.98 [0.97, 0.98]} & \textbf{0.95 [0.94, 0.96]} \\
\bottomrule\end{tabular}}\end{table}

\begin{table}[!t]\centering\footnotesize\setlength{\tabcolsep}{3pt}
\caption{Precipitation pilot competition: Brier scores for exceeding 10 and 25\,mm relative to the reference (censored-Gaussian kriging (v)), as in Table~\ref{tab:cerra}. Bold: the interval excludes one (decided at full precision, so a limit printed as 1.000 can be bold); a limit that rounds to 1.00 is given to three decimals.}\label{tab:cerrabrier}
\resizebox{\textwidth}{!}{\begin{tabular}{lllllll}\toprule
 & \multicolumn{3}{c}{2\% observed} & \multicolumn{3}{c}{0.5\% observed} \\ \cmidrule(lr){2-4}\cmidrule(lr){5-7}
Method & 3 years & 10 years & 30 years & 3 years & 10 years & 30 years \\ \midrule
\multicolumn{7}{l}{\emph{Brier score, 10\,mm}} \\
(i) Kriging, day alone & \textbf{1.16 [1.15, 1.18]} & \textbf{1.23 [1.22, 1.25]} & \textbf{1.29 [1.28, 1.30]} & \textbf{1.27 [1.25, 1.29]} & \textbf{1.29 [1.28, 1.31]} & \textbf{1.32 [1.30, 1.34]} \\
(ii) Archive kriging & \textbf{1.02 [1.01, 1.03]} & \textbf{0.98 [0.97, 0.99]} & \textbf{0.97 [0.96, 0.98]} & \textbf{1.04 [1.03, 1.05]} & \textbf{1.02 [1.01, 1.03]} & \textbf{1.02 [1.01, 1.03]} \\
(iv) Log-scale archive kriging & \textbf{1.01 [1.01, 1.02]} & 1.00 [0.99, 1.001] & \textbf{0.99 [0.98, 0.99]} & \textbf{1.02 [1.01, 1.02]} & \textbf{1.01 [1.01, 1.02]} & \textbf{1.01 [1.01, 1.02]} \\
(v) Censored-Gaussian kriging & reference & reference & reference & reference & reference & reference \\
(vi) TabICL & \textbf{0.95 [0.94, 0.96]} & \textbf{0.99 [0.98, 0.996]} & \textbf{1.02 [1.01, 1.02]} & \textbf{1.07 [1.06, 1.08]} & \textbf{1.08 [1.07, 1.09]} & \textbf{1.10 [1.09, 1.11]} \\
(vii) TabICL + kriging & \textbf{0.94 [0.93, 0.94]} & \textbf{0.96 [0.95, 0.96]} & \textbf{0.97 [0.96, 0.98]} & \textbf{1.10 [1.09, 1.11]} & \textbf{1.11 [1.10, 1.12]} & \textbf{1.13 [1.12, 1.15]} \\
(viii) Diffusion & \textbf{0.96 [0.95, 0.97]} & \textbf{0.89 [0.89, 0.90]} & \textbf{0.87 [0.86, 0.88]} & 1.01 [0.997, 1.01] & \textbf{0.97 [0.96, 0.98]} & \textbf{0.95 [0.94, 0.95]} \\
\midrule
\multicolumn{7}{l}{\emph{Brier score, 25\,mm}} \\
(i) Kriging, day alone & \textbf{1.11 [1.10, 1.13]} & \textbf{1.18 [1.17, 1.20]} & \textbf{1.24 [1.23, 1.26]} & \textbf{1.20 [1.18, 1.23]} & \textbf{1.22 [1.20, 1.25]} & \textbf{1.26 [1.24, 1.28]} \\
(ii) Archive kriging & \textbf{0.98 [0.97, 0.99]} & \textbf{0.96 [0.95, 0.97]} & \textbf{0.96 [0.95, 0.97]} & \textbf{1.03 [1.02, 1.04]} & \textbf{1.02 [1.002, 1.03]} & \textbf{1.02 [1.005, 1.03]} \\
(iv) Log-scale archive kriging & \textbf{1.06 [1.05, 1.07]} & \textbf{1.03 [1.02, 1.03]} & \textbf{1.02 [1.01, 1.02]} & \textbf{1.02 [1.02, 1.03]} & \textbf{1.01 [1.01, 1.02]} & \textbf{1.01 [1.005, 1.02]} \\
(v) Censored-Gaussian kriging & reference & reference & reference & reference & reference & reference \\
(vi) TabICL & \textbf{0.93 [0.92, 0.94]} & \textbf{0.97 [0.96, 0.98]} & 1.01 [0.996, 1.01] & \textbf{1.14 [1.11, 1.16]} & \textbf{1.14 [1.12, 1.16]} & \textbf{1.16 [1.14, 1.19]} \\
(vii) TabICL + kriging & \textbf{0.92 [0.91, 0.93]} & \textbf{0.94 [0.93, 0.95]} & \textbf{0.96 [0.95, 0.97]} & \textbf{1.15 [1.13, 1.17]} & \textbf{1.15 [1.13, 1.18]} & \textbf{1.18 [1.16, 1.21]} \\
(viii) Diffusion & \textbf{0.94 [0.94, 0.95]} & \textbf{0.87 [0.87, 0.88]} & \textbf{0.85 [0.84, 0.86]} & 1.00 [0.98, 1.01] & \textbf{0.97 [0.96, 0.98]} & \textbf{0.94 [0.93, 0.95]} \\
\bottomrule\end{tabular}}\end{table}

\begin{table}[!t]\centering\footnotesize\setlength{\tabcolsep}{3pt}
\caption{Precipitation pilot competition, zero-shot: models that see no Alpine history, relative to kriging fitted to the day's observations alone (i), with the intervals of Table~\ref{tab:cerra}; the last row is the same diffusion model trained on the Alps' own 30 years, also without the input (ii). Bold: the interval excludes one (decided at full precision, so a limit printed as 1.000 can be bold); a limit that rounds to 1.00 is given to three decimals.}\label{tab:cerrazero}
\resizebox{\textwidth}{!}{\begin{tabular}{lllllllll}\toprule
 & \multicolumn{4}{c}{2\% observed} & \multicolumn{4}{c}{0.5\% observed} \\ \cmidrule(lr){2-5}\cmidrule(lr){6-9}
Method & RMSE & CRPS & Brier 10 & Brier 25 & RMSE & CRPS & Brier 10 & Brier 25 \\ \midrule
Kriging, Carpathian covariance & \textbf{1.01 [1.01, 1.01]} & 1.00 [0.997, 1.002] & \textbf{1.01 [1.01, 1.02]} & \textbf{1.02 [1.01, 1.02]} & \textbf{0.98 [0.97, 0.98]} & \textbf{0.97 [0.97, 0.98]} & \textbf{0.98 [0.97, 0.98]} & \textbf{0.98 [0.97, 0.99]} \\
Kriging, Pyrenean covariance & \textbf{1.01 [1.003, 1.01]} & \textbf{1.00 [0.99, 0.998]} & \textbf{1.01 [1.004, 1.01]} & \textbf{1.01 [1.01, 1.02]} & \textbf{0.97 [0.97, 0.98]} & \textbf{0.97 [0.97, 0.97]} & \textbf{0.98 [0.97, 0.98]} & \textbf{0.97 [0.97, 0.98]} \\
TabICL, day alone & \textbf{1.03 [1.02, 1.04]} & \textbf{0.92 [0.91, 0.92]} & \textbf{0.94 [0.94, 0.95]} & 1.00 [0.99, 1.01] & \textbf{1.08 [1.07, 1.10]} & 1.00 [0.99, 1.01] & \textbf{0.98 [0.96, 0.99]} & \textbf{1.09 [1.08, 1.11]} \\
Diffusion trained on the Carpathians & \textbf{1.13 [1.11, 1.16]} & \textbf{0.98 [0.97, 0.996]} & \textbf{1.03 [1.02, 1.04]} & \textbf{1.11 [1.09, 1.12]} & \textbf{1.11 [1.09, 1.13]} & 1.00 [0.98, 1.02] & \textbf{1.04 [1.02, 1.06]} & \textbf{1.09 [1.06, 1.11]} \\
Diffusion trained on the Pyrenees & \textbf{1.06 [1.05, 1.07]} & \textbf{0.93 [0.92, 0.94]} & \textbf{0.98 [0.97, 0.99]} & \textbf{1.06 [1.05, 1.07]} & \textbf{1.05 [1.03, 1.06]} & \textbf{0.94 [0.93, 0.95]} & \textbf{0.97 [0.96, 0.99]} & \textbf{1.05 [1.03, 1.07]} \\
Diffusion trained on the Alps (30 years) & \textbf{0.77 [0.76, 0.78]} & \textbf{0.66 [0.65, 0.66]} & \textbf{0.70 [0.69, 0.70]} & \textbf{0.71 [0.70, 0.72]} & \textbf{0.83 [0.81, 0.84]} & \textbf{0.72 [0.71, 0.73]} & \textbf{0.74 [0.73, 0.75]} & \textbf{0.78 [0.76, 0.79]} \\
\bottomrule\end{tabular}}\end{table}

\FloatBarrier
\bibliographystyle{sn-basic}
\bibliography{refs}